\documentclass[12pt]{article}
\usepackage{amssymb,graphicx,epsfig,psfig,color}
\renewcommand{\theequation}{\thesection.\arabic{equation}}

\begin{document}

\begin{flushright}
{\tt hep-th/0601206}
\end{flushright}

\vspace{5mm}

\begin{center}
{{{\Large \bf Tachyon Kinks in Boundary String Field Theory}}\\[12mm]
{Chanju Kim}\\[1mm]
{\it Department of Physics, Ewha Womans University,
Seoul 120-750, Korea}\\
{\tt cjkim@ewha.ac.kr}\\[6mm]
{Yoonbai Kim~~and~~O-Kab Kwon}\\[1mm]
{\it BK21 Physics Research Division and Institute of
Basic Science,\\
Sungkyunkwan University, Suwon 440-746, Korea}\\
{\tt yoonbai@skku.edu~~~okab@skku.edu}\\[6mm]
{Ho-Ung Yee}\\[1mm]
{\it Korea Institute for Advanced Study,\\
207-43 Cheongryangri 2-dong Dongdaemun-gu, Seoul 130-722, Korea}\\
{\tt ho-ung.yee@kias.re.kr}
}
\end{center}
\vspace{10mm}

\begin{abstract}
We study tachyon kinks with and without electromagnetic fields
in the context of boundary string field theory.
For the case of pure tachyon only an array of kink-antikink is obtained.
In the presence of electromagnetic coupling, all possible static
codimension-one soliton solutions such as array of kink-antikink,
single topological BPS kink, bounce, half kink,
as well as nonBPS topological kink are found, and their properties
including the interpretation as branes are analyzed in detail.
Spectrum of the obtained kinks coincides with that of
Dirac-Born-Infeld type effective theory.
\end{abstract}


\newpage

\setcounter{equation}{0}
\section{Introduction}

Study on the unstable systems of D-branes
has been an attractive subject in string theory; for example, an unstable
D-brane or
D${\bar {\rm D}}$~\cite{Sen:2004nf}.
An intriguing issue among various  questions is  the pattern of remnant
after the unstable D$p$-brane or D$p{\bar {\rm D}}p$ decays completely.
Since perturbative open string degrees of freedom should disappear
in the absence
of the original branes,
the remained fossils can be either closed strings or
nonperturbative objects in open string theory.

When nonperturbative open string degrees of freedom are considered
in a form of
solitonic objects, a lesson from quantum field theories dealing with
spontaneous symmetry breaking is that we should employ a scheme of
semi-classical approximation. There are several appropriate languages
describing open string tachyons and the solitonic
lower dimensional branes, including
boundary conformal field theory
(BCFT)~\cite{Sen:2002nu,Sen:2003bc},
boundary string field theory (BSFT or background-independent string field
theory)~\cite{Gerasimov:2000zp,Kutasov:2000aq,Sugimoto:2002fp,Lambert:2003zr},
effective field theory (EFT)~\cite{Minahan:2000tf,Sen:2003tm,Kim:2003in},
and noncommutative field
theory (NCFT) in the presence of fundamental strings~\cite{Gopakumar:2000zd}.
If we want to take into account string off-shell contributions,
 BSFT  fits its purpose. However, most of the previous of studies
 have been made
in the context of Dirac-Born-Infeld (DBI) type tachyon EFT with runaway
potential, avoiding the complicated  kinetic term of the BSFT
action~~\cite{Gerasimov:2000zp,Kutasov:2000aq,Kraus:2000nj}.

In BSFT, decent relations for the codimension-one and -two
branes were obtained in exact form from the energy density difference
between the false and true
vacua~\cite{Gerasimov:2000zp,Kutasov:2000aq},
but the corresponding smooth solitonic solutions of the equations of motion
have not been analyzed before in BSFT.
On the other hand, the tachyon
kinks~\cite{Sen:2003tm,Lambert:2003zr,Kim:2003in,Brax:2003rs,Copeland:2003df},
tubes~\cite{Kim:2003uc}, and
vortices~\cite{Sen:2003tm,Kim:2005tw} have been found
by examining the equations of motion from the DBI type EFT.
Specifically, for the case of tachyon kinks, pure tachyon can support only an
array of kink-antikink~\cite{Lambert:2003zr,Kim:2003in}.
In the presence of the DBI electromagnetic fields, the spectrum of
the kinks solutions becomes
richer; they include array of kink-antikink, single topological BPS kink,
bounce, half-kink, as well as topological nonBPS
kink~\cite{Kim:2003in,Kim:2003ma}.
Moreover, with the $1/\cosh$-type tachyon potential,
analytic solutions are available~\cite{Kim:2003in}
and their tachyon profiles can be mapped to those
in BCFT by Ref.~\cite{Kutasov:2003er}.
These kink solutions are also reproduced in
NCFT~\cite{Banerjee:2004cw}.

In this paper, we study the effective action from BSFT
with worldsheet supersymmetry and find all the aforementioned tachyon kinks
as solutions of classical equations of motion. Moreover,
the solution of  single
topological BPS kink with critical DBI electromagnetic field
is given in a closed form, so that
the exact solution may be useful in studying the
relation between BCFT and BSFT.
The species of the tachyon
kinks coincide with those in the DBI type EFT,  but they show
slight difference in some minor details such as shape of energy density.
This seems to suggest a need of further study in BCFT. Since only array of
kink-antikink and single topological BPS kink were discovered in  BCFT
with electromagnetic coupling~\cite{Sen:2003bc},
it would be intriguing if we reproduce other three
kinks (bounce, half kink, topological nonBPS kink) in the context of
BCFT~\cite{KKKY}.
The former two kink configurations can easily be
interpreted as array of D$(p-1){\bar {\rm D}}(p-1)$ or
D$(p-1)$F1${\bar {\rm D}}(p-1)$F1 and single thick D$(p-1)$ or
D$(p-1)$F1. Though we discuss some interpretations as brane
configurations, the other three kink configurations await proper
understanding in string theory.

In section 2 we  briefly review a derivation of the super BSFT
action with a real
tachyon and DBI type electromagnetic field, and derive
its classical equations of motion.
In sections 3 and 4, we consider the tachyon without and with DBI type
electromagnetic field respectively,
and find all possible static kink solutions from the equations
of motion. The obtained spectrum of the kinks coincides with that from
DBI type EFT and NCFT.
Section 5 contains our conclusions with brief outlook for further study.
Appendix A includes detailed calculations of array of kink-antikink,
and appendix B deals with comparison of the tachyon kinks
between super and bosonic BSFTs.

\setcounter{equation}{0}
\section{BSFT Setup}

In this section we briefly recapitulate derivation of the BSFT action of
the real tachyon and gauge field on an unstable D$p$-brane
in superstring theory.
According to the BV-like formulation of BSFT for
open string~\cite{Witten:1992qy},
the off-shell BSFT action is identical to the disc partition
function~\cite{Marino:2001qc} including gauge
fields~\cite{Tseytlin:1987ww} and
tachyon~\cite{Kutasov:2000aq,Kraus:2000nj};
\begin{eqnarray}
S=Z.
\end{eqnarray}
For an unstable D$p$-brane, the partition function is given by
\begin{eqnarray}
Z=\int [{\cal D}X^{\mu}][{\cal D}\psi^{\mu}][{\cal D}{\bar \psi}^{\mu}]
\exp\left[-(S_{{\rm bulk}}+S_{{\rm bdry}})\right],
\end{eqnarray}
where
the bulk action $S_{{\rm bulk}}$ and the boundary action $S_{{\rm bdry}}$ are
\begin{eqnarray}
S_{{\rm bulk}}&=& \frac{1}{4\pi}\int_{\Sigma}d^{2}z\left(
\partial X^{\mu}{\bar \partial}X_{\mu}
+\psi^{\mu}{\bar \partial}\psi_{\mu}+
{\bar \psi}^{\mu}\partial {\bar \psi}_{\mu}\right),\\
S_{{\rm bdry}}&=& \int_{\partial\Sigma}\frac{d\tau}{2\pi}
\left[\frac{1}{4}T^{2}-\frac{1}{2}\psi^{\mu}\partial_{\mu}T
\partial_{\tau}^{-1}(\psi^{\nu}\partial_{\nu}T)
+\frac{i}{2} F_{\mu\nu}\psi^{\mu}\psi^{\nu}-\frac{i}{2}A_{\mu}
\frac{dX^{\mu}}{d\tau}
\right].
\end{eqnarray}
Here $\Sigma$ denotes a disc with unit radius in the Euclidean worldsheet
with complex coordinates $(z,{\bar z})$, and its boundary is $\partial\Sigma$
with coordinate $\tau$ ($0\le \tau \le 2\pi$).
Note that we set $\alpha'=2$ and contributions from ghost sector are decoupled.

Let us consider a relevant linear profile of the tachyon
\begin{eqnarray}
T(X)= u_{\mu}X^{\mu},
\end{eqnarray}
and a constant electromagnetic field strength $F_{\mu\nu}$ with symmetric gauge
$A_{\mu}=-F_{\mu\nu}X^{\nu}/2$.
Performing functional integration of the worldsheet fields, we
obtain an action in a closed form
\begin{eqnarray}
S=Z&=&Z_{{\rm F}}Z_{{\rm B}}\\
&\sim&\int d^{d}x\, e^{-\frac{1}{4}T^{2}}
\frac{\prod_{r=\frac{1}{2}}^{\infty}\det(\eta_{\mu\nu}
+\frac{1}{r}\partial_{\mu}T\partial_{\nu}T+F_{\mu\nu})}{
\prod_{n=1}^{\infty}\det(n\eta_{\mu\nu}
+\partial_{\mu}T\partial_{\nu}T+ nF_{\mu\nu})},
\label{acfb}
\end{eqnarray}
where $Z_{{\rm F}}$ and $Z_{{\rm B}}$ are fermionic and bosonic
parts of the worldsheet partition function, respectively.
The action (\ref{acfb}) is written as a divergent infinite product,
and through zeta function regularization, we obtain a well-defined finite
action of the tachyon and the DBI electromagnetic field. The result is
\begin{eqnarray}\label{yac}
S(T,A_{\mu})= -{\cal T}_{p}\int d^{d}x \, V(T)
\sqrt{-\det (\eta_{\mu\nu}+F_{\mu\nu})}\,
{\cal F}(y),
\end{eqnarray}
where the potential is
\begin{eqnarray}
V(T)&=&e^{-\frac{1}{4}T^{2}}\label{pot},
\end{eqnarray}
and the kinetic piece is
\begin{eqnarray}
{\cal F}(y)&=&\frac{4^{y}y\, \Gamma(y)^{2}}{2\Gamma(2y)}
=2^{2y-1}y B(y,y)
=y\int_{-1}^{1}dx \, (1-x^{2})^{y-1} \label{y}
\end{eqnarray}
with $y=\left(\frac{1}{\eta+F}\right)^{\mu\nu}\partial_{\mu}T\partial_{\nu}T$.
We also fix the overall normalization by the tension ${\cal T}_{p}$
of the unstable D$p$-brane.

It is straightforward to derive the equations of motion.
For tachyon field, it reads as
\begin{eqnarray}\label{teq}
\partial_{\mu}\left[V\sqrt{-\det(\eta+F)}\,{ d{\cal F}\over dy}
\left(\frac{1}{\eta+F}\right)^{\mu\nu}_{{\rm S}}\partial_{\nu}T \right]
=\frac{1}{2}\sqrt{-\det(\eta+F)}\, {\cal F}\frac{dV}{dT},
\end{eqnarray}
while for gauge field, it can be written as
\begin{eqnarray}\label{geq}
\partial_{\mu}\Pi^{\mu\nu}=0
\end{eqnarray}
with the antisymmetric displacement tensor $\Pi^{\mu\nu}$ given by
\begin{eqnarray}
\Pi^{\mu\nu}&\equiv&\frac{\delta S}{\delta \partial_{\mu}A_{\nu}}
={\cal T}_{p}V\sqrt{-\det(\eta+F)}\left[
\left(\frac{1}{\eta+F}\right)^{\mu\nu}_{{\rm A}}{\cal F} +
\right.  \\
&&\left.
2\left(\left(\frac{1}{\eta+F}\right)^{\rho\mu}_{{\rm A}}
\left(\frac{1}{\eta+F}\right)^{\sigma\nu}_{{\rm S}}
-\left(\frac{1}{\eta+F}\right)^{\rho\nu}_{{\rm A}}
\left(\frac{1}{\eta+F}\right)^{\sigma\mu}_{{\rm S}}
\right){ d{\cal F}\over dy}\partial_{\rho}T\partial_{\sigma}T\right].\nonumber
\label{pmn}
\end{eqnarray}
For the tachyon equation (\ref{teq}), it is possible to consider
conservation of energy-momentum instead,
\begin{equation}\label{emc}
\partial_{\mu}T^{\mu\nu}=0,
\end{equation}
where the energy-momentum tensor read from the above action is
\begin{eqnarray}\label{tmn}
T^{\mu\nu}&=& -{\cal T}_{p}V(T)\sqrt{-\det(\eta+F)}
\left\{ {\cal F}(y)\left(\frac{1}{\eta+F}\right)^{\mu\nu}_{{\rm S}}
\right. \\
&&\left. -{d{\cal F}\over dy}\left[\left(\frac{1}{\eta+F}\right)^{\alpha\mu}
\left(\frac{1}{\eta+F}\right)^{\nu\beta}
+\left(\frac{1}{\eta+F}\right)^{\alpha\nu}
\left(\frac{1}{\eta+F}\right)^{\mu\beta}
\right]\partial_{\alpha}T\partial_{\beta}T\right\}.
\nonumber
\end{eqnarray}

We are interested in static kink configurations satisfying
the above equations of motion
(\ref{teq})--(\ref{geq}), and we will interpret the obtained solitons as
codimension one, BPS composites of D$(p-1)$-branes and fundamental
strings.
The case of pure tachyon field with vanishing electromagnetic field is
considered in section 3, and we analyze the case with electromagnetic
field turned on in section 4.

\setcounter{equation}{0}
\section{Array of D$(p-1){\bar {\bf D}}(p-1)$ as Tachyon Kink}
We begin this section with turning off the gauge field $F_{\mu\nu}=0$
and considering a simple case of the pure tachyon field.
Then the equation of the gauge field (\ref{geq}) is trivially satisfied.
To construct flat codimension-one soliton
configurations by examining the classical tachyon equation
(\ref{teq}), we naturally introduce an ansatz
\begin{equation}\label{an}
T=T(x),
\end{equation}
where $x$ is one of spatial coordinates.

For the case of pure tachyon, substitution of the ansatz (\ref{an})
makes the equations (\ref{emc}) simple, i.e.,
the only equation we have to solve is
that the parallel component of the pressure should be constant;
${d T^{11} \over dx}\equiv (T^{11})'=0$,
where
\begin{equation}\label{t11}
T^{11}=-{\cal T}_p V(T)G(y)=-{\cal T}_p \,e^{-{1 \over 4} T^2}G(y),
\qquad y=T'^{2}\ge 0,
\end{equation}
and the function $G(y)$ defined by
\begin{equation}\label{G}
G(y)={\cal F}(y)-2y\dot{\cal F}(y),\qquad
{\dot {\cal F}}(y)=\frac{d{\cal F}}{dy},
\end{equation}
is monotonically-decreasing from $G(0)=1$ to $G(\infty)=0$.
Introducing functions
\begin{eqnarray}
K(y)&=&\frac{1}{G(y)^{2}}-1,
\label{ki}\\
U(T)&=&-\left(\frac{{\cal T}_{p}V(T)}{T^{11}}\right)^{2}
=-\left(\frac{{\cal T}_{p}}{T^{11}}\right)^{2}e^{-\frac12 T^2},
\label{po}
\end{eqnarray}
we rewrite the equation (\ref{t11}) as
\begin{equation}\label{ene}
{\cal E}=K(T'^{2})+U(T).
\end{equation}
where ${\cal E}=-1$,
$K(y)$ is a positive, monotonically-increasing function with
the minimum ${\rm min}(K)=0$ at $y=0$,
and the profile of $U(T)$ has ${\rm min}(U)=U(0)=-({\cal
T}_{p}/T^{11})^{2}$ and is monotonically-increasing to ${\rm
max}(U)=U(T=\pm\infty)=0$.
Suppose we consider one-dimensional motion of a hypothetical particle
at a position $T$ and time $x$ with conserved mechanical energy ${\cal E}$
and velocity $T'$.
Then the position-dependent function $U(T)$ may be identified as a kind of
``potential'' and the velocity-dependent function $K(T'^{2})$ as a kind of
``kinetic energy''.
The reason for choosing this form of kinetic term $K(T'^2)$ (\ref{ki}) is that
it becomes quadratic in $T'$ both in $|T'|\to 0$ and $|T'|\to \infty$ limits:
\begin{equation}\label{aK}
K(T'^2)\to \Bigg\{
\begin{array}{ccc}  {16\over \pi}\,
T'^2 +\left(-1- {4\over \pi}\right)+{\cal O}(|T'|^{-2}),
&& |T'|\to \infty \\
(4\log{2})\, T'^2 + {\cal O}(|T'|^4),&  & |T'| \to 0
\end{array}
\end{equation}
so that it mimics a right kind of kinetic term.
\begin{figure}[ht]
\begin{center}
\scalebox{0.6}[0.6]{\includegraphics{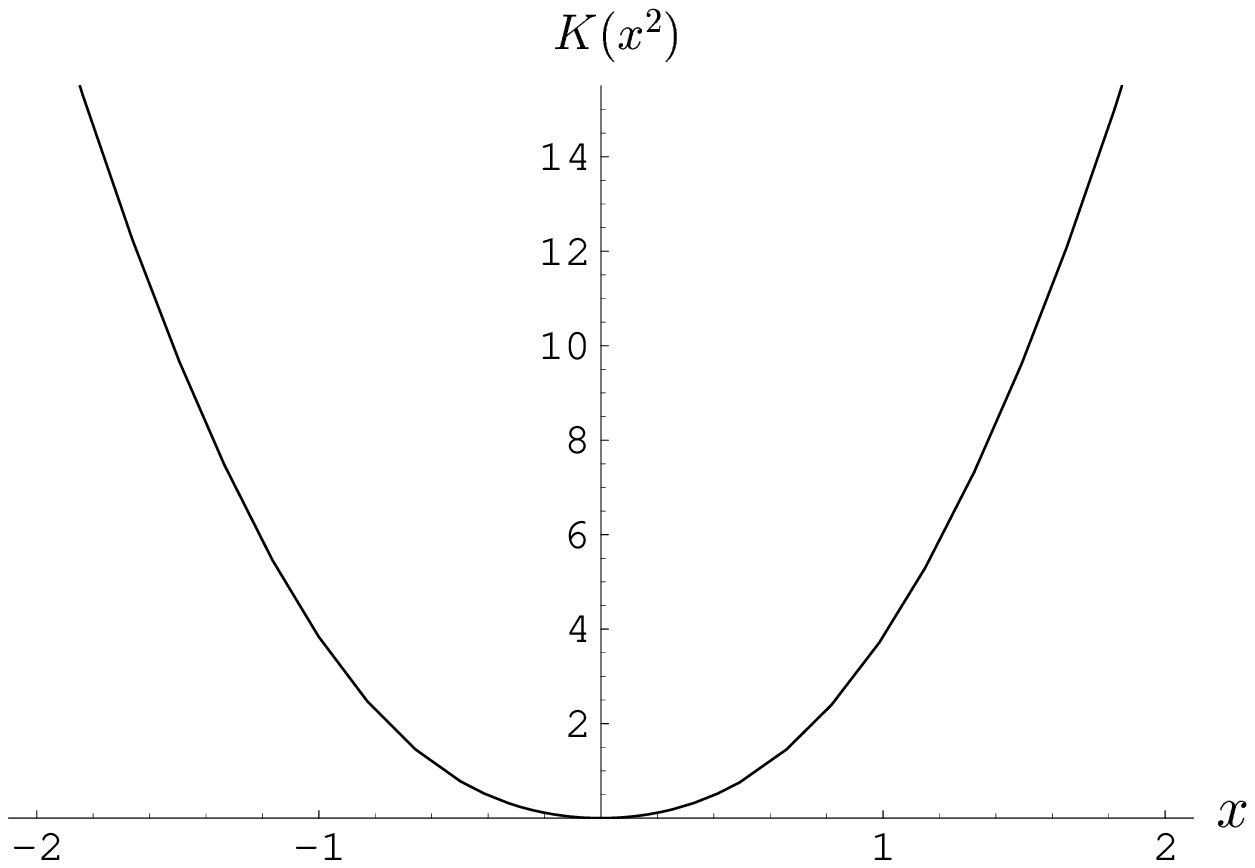}
\includegraphics{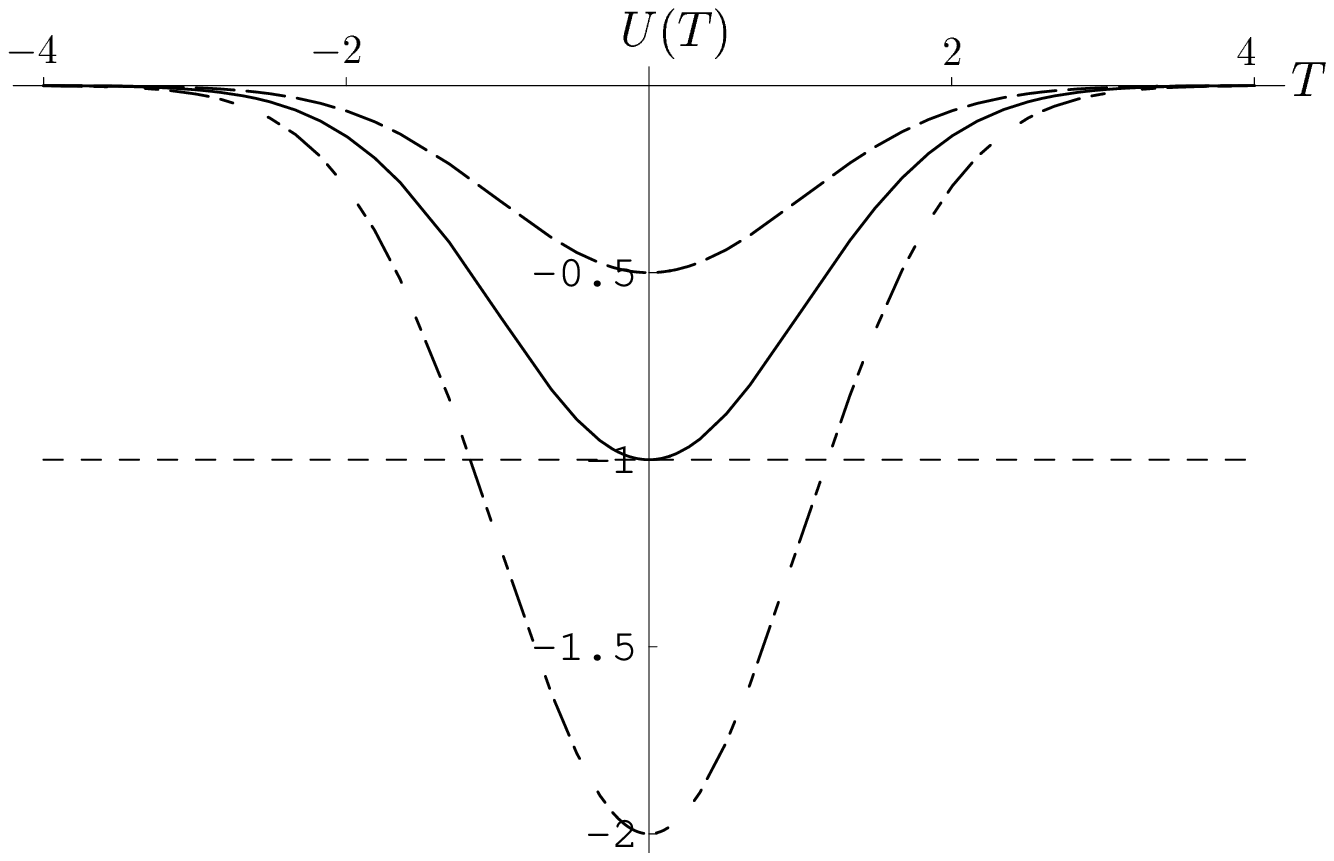}}
\par
\vskip-2.0cm{}
\end{center}
\caption{\small The graphs of $K(T'^{2})$(left) and
$U(T)$(right).  For $U(T)$, three cases of ${{\cal T}_p / (-T^{11}})
=1/ \sqrt{2}~\mbox{(dashed curve)},\;1~\mbox{(solid curve)},\;
\sqrt{2}~\mbox{(dotted-dashed curve)}$
from top to bottom are shown. ${\cal E}=1$ is denoted by dotted line.}
\label{fig1}
\end{figure}
Fig.~\ref{fig1} shows the graphs of $K(T'^{2})$ and $U(T)$.

\begin{figure}[ht]
\begin{center}
\scalebox{1}[1]{\includegraphics{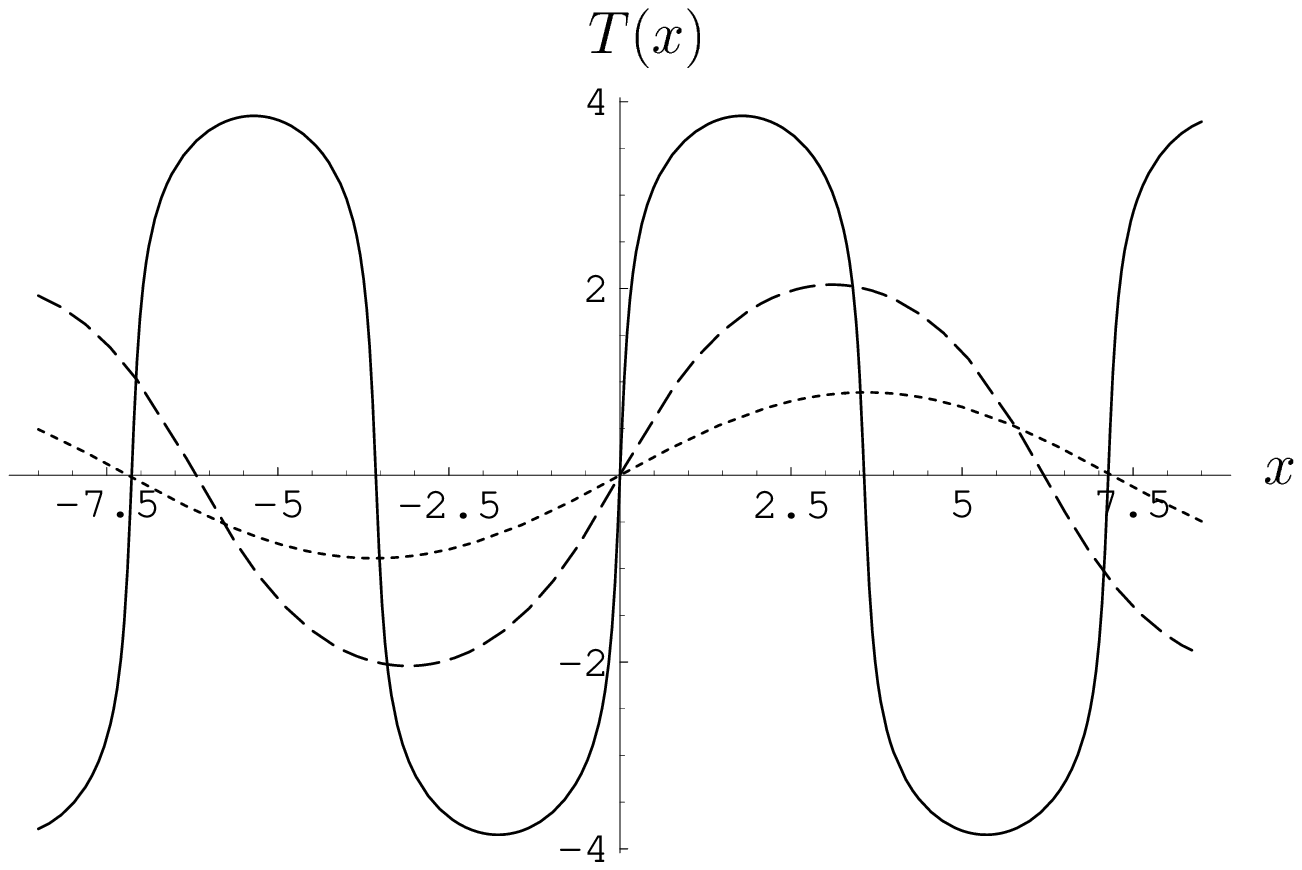}}
\par
\vskip-2.0cm{}
\end{center}
\caption{\small Profiles of array of kink-antikink for the cases  $T_{\rm
max}=1,2,3.8$ from bottom to top are shown. It is clear that
their width is decreasing as $T_{\rm max}$ increases.}
\label{fig2}
\end{figure}

From this, we see that there always
exists a nontrivial {\it oscillating}\/ solution $T(x)$ when
$U(0)=-({\cal T}_{p}/T^{11})^{2}< -1$.
At the maximum of $|T(x)|$, we have $T'=0$ and
(\ref{ene}) gives the maximum value,
\begin{equation}\label{max}
T_{\rm max}=2\sqrt{\ln|{\cal T}_{p}/T^{11}|}\, .
\end{equation}
Since no exact solution is obtained in closed form, we plot
numerical solutions in Fig.~\ref{fig2}. This solution can be
interpreted as an array of kink-antikink pairs.

\begin{figure}[ht]
\begin{center}
\scalebox{1}[1]{\includegraphics{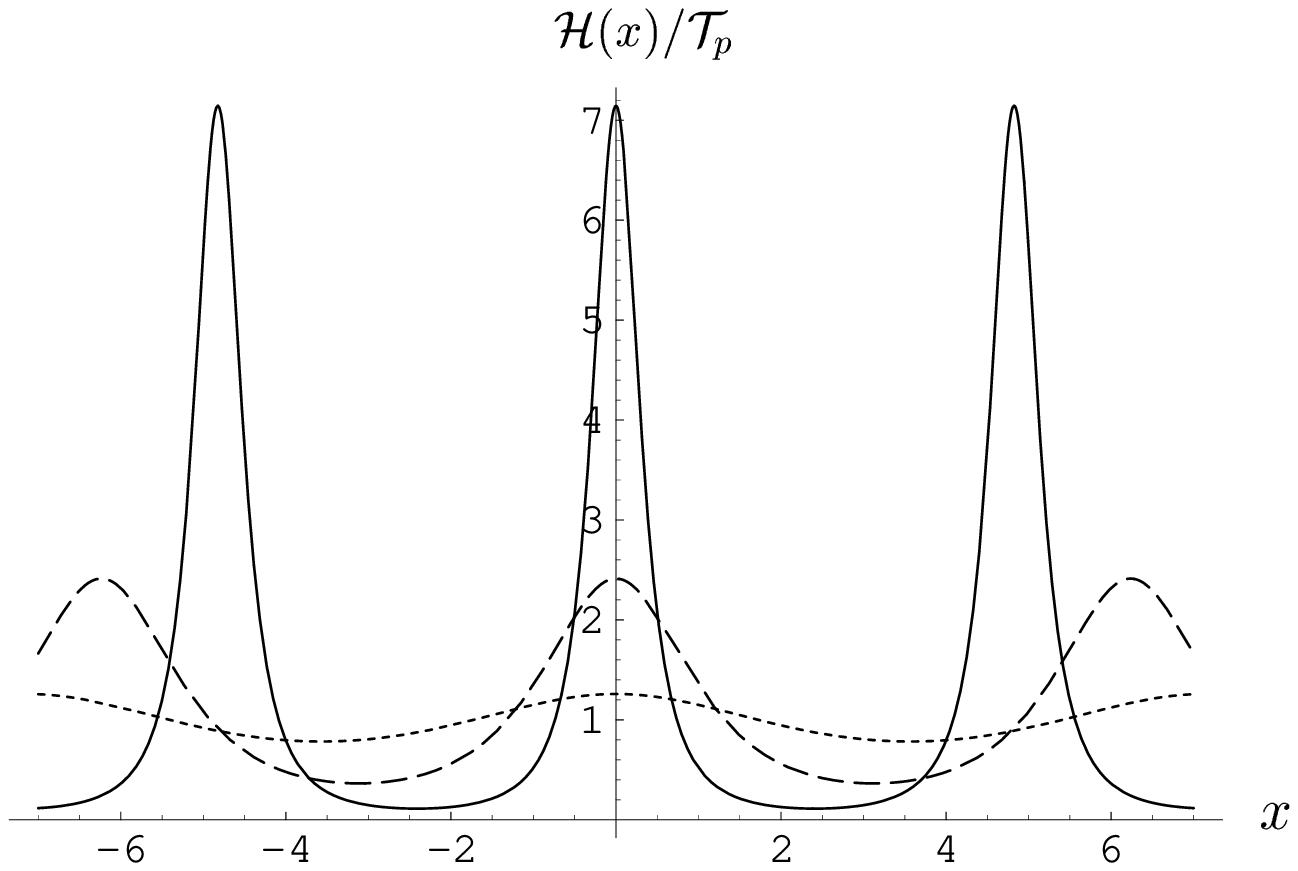}}
\par
\vskip-2.0cm{}
\end{center}
\caption{\small Energy density profiles of single kink (antikink)
for the cases  $T_{\rm
max}=1,2,3$ from bottom to top are shown. It is clear that
their width is decreasing as $T_{\rm max}$ increases.}
\label{fig3}
\end{figure}

We also find from the ansatz (\ref{an}) that
the momentum density vanishes, $T^{01}=0$, and the energy density is
the only quantity involving local physical character
\begin{equation}\label{ede}
T^{00}={\cal T}_p\, V(T){\cal F}(T'^{2})={\cal T}_p \,e^{-{1 \over 4} T^2}\,
{\cal F}(T'^{2}).
\end{equation}
For the obtained solution, the energy density of a
single kink (antikink) is shown in Fig.~\ref{fig3} for several $T_{\max}$.
We see that the energy is obviously accumulated at the site of kink (antikink).

Note that, in Fig.~\ref{fig2}, a kink and its adjacent antikink gets closer
as $T_{\max}$ becomes larger (or, equivalently,  $-T^{11}$ becomes smaller).
This is in contrast with the array solution in DBI-type EFT with
$1/\cosh$ potential where the distance is constant regardless of the
value of $T^{11}$ \cite{Kim:2003in}. Therefore it is interesting to
find how the distance changes as we vary the negative pressure
$-T^{11}$ which is the only parameter governing the array solution.

Let us first consider the limit $-T^{11}\rightarrow{\cal T}_p$ which
corresponds to $T_{\max}\rightarrow0$, i.e., the solution represents
a small fluctuation around vacuum. Then we have a harmonic
oscillator solution
\begin{equation}
T(x)\,\,\approx\,\, T_{\rm max}\,\sin\left({2\pi\over \xi_0} \,x\right),
\end{equation}
where the width $\xi_0$ is equal to $2\pi\sqrt{2\log{2}}\approx 7.39$.

In the opposite limit of $T_{\max} \rightarrow \infty$ or
$T^{11} \rightarrow 0$, complicated nature of the ``kinetic'' term $K(T'^2)$
comes into play and a careful investigation is needed.
Since the analysis is rather technical, here we will only
sketch the argument. Details can be found in Appendix A.
First observe from Fig.~\ref{fig2} that the array solution can be
divided roughly into three regions, namely the region around
(anti)kink sites where the slope of the tachyon profile is steep,
plateau region where there is almost no change in $T(x)$,
and the transition region which connects the other two regions.
From the figure it is easy to guess that the width of the (anti)kink region
goes to zero in $T^{11} \rightarrow 0$ limit. Then it remains to estimate the
distance of the other two regions. According to the analysis of Appendix A,
the result is that
actually the width of the other regions also vanishes as $T^{11}$ goes to zero.
This behavior is essentially related
with large $T$ behavior of the potential which vanishes faster than
$e^{-T}$ in this theory. This is in accordance with the results in DBI type
EFT \cite{Brax:2003rs,Copeland:2003df}.

Finally, we come to the question of energy density in large $T_{\max}$ limit.
As seen in Fig.~\ref{fig3}, the energy stored in the plateau region and
the transition region is negligible. (This is shown more rigorously in
Appendix A.) Since $T'^2$ becomes very large in (anti)kink region,
we can approximate ${\cal F}(T'^2) \approx \sqrt\pi \, T'$ and the energy
integral for a kink becomes
\begin{equation}
\int \, T^{00}\,dx \,\,\approx\,\,\sqrt{\pi}\, {\cal T}_p
\int\,e^{-{1\over 4}T^2}\,T'\,dx\,\,=\,\,\sqrt{\pi}\, {\cal T}_p
\int\,e^{-{1\over 4}T^2}\,dT.
\end{equation}
This becomes exact as $T_{\rm max}\to \infty$ ($T^{11}\to 0$) and hence
\begin{equation} \label{puret}
{\cal T}_{p-1}=\sqrt{\pi}\, {\cal T}_p\,
\int_{-\infty}^{\infty}\,\,e^{-{1\over 4}T^2}\,dT
=2\pi\,{\cal T}_p =(2\pi\sqrt{\alpha'} )\,{{\cal T}_p\over \sqrt{2}},
\end{equation}
reproducing the tension of BPS D$(p-1)$ brane.

To summarize, pressureless limit leads to the BPS kink which interpolates
two tachyon vacua $T=\pm T_{\max} = \pm\infty$ with
right tension  and the obtained static regular configuration describes
an array of kink-antikink, which is the same as the soliton spectrum
in DBI type EFT and BCFT.
In the context of superstring theory, it is interpreted as
an array of D$(p-1){\bar {\rm D}}(p-1)$.

\setcounter{equation}{0}
\section{Codimension-one Branes from Tachyon and Electromagnetic Field}

If we turn on DBI electromagnetic field $F_{\mu\nu}$
on the unstable D$p$-brane, the analyses based on DBI type
EFT~\cite{Kim:2003in,Kim:2003ma}, BCFT~\cite{Sen:2003bc},
and NCFT~\cite{Banerjee:2004cw}
showed that spectrum of the regular kinks and their composites becomes rich.
To be specific, the ansatz for codimension-one objects,
\begin{equation}\label{fmn}
F_{\mu\nu}=F_{\mu\nu}(x),
\end{equation}
was made for DBI electromagnetic field in EFT and NCFT,
but the classical equations dictate it to be
constant which is consistent with the assumption in BCFT.
Examining the remained equations, they obtained various kinds of static
kink configurations.
For $p=1$, array of kink-antikink, single topological BPS kink,
and topological non-BPS kink have been found in both EFT and
NCFT but only first two of them were identified in BCFT.
For $p\ge 2$, in EFT and NCFT, three more types of kinks
were found in addition to
the above three, i.e., they are bounce, half kink, and hybrid of two
half kinks (we call the hybrid as another topological nonBPS kink
in what follows),
however they have not been identified in BCFT.
Furthermore, proper interpretation of the kink configurations which
are not found in the context of BCFT has not been made, yet.

Since BSFT is valid for description of off-shell physics of the string theory
and its classical limit arrives at a BCFT,
it is an appropriate scheme to reconcile this discrepancy.
In the following two subsections, we investigate static kink solutions in BSFT
and interpret the obtained solutions in terms of D-branes and their
composites with fundamental strings.

\subsection{$p=1$}
When we have an unstable D1-brane, there exist a tachyon field and
an electric field component $E=-F_{01}$. For static objects, the
equations we have to deal with reduce to the conservation of energy-momentum
(\ref{emc}), $(T^{11})'=0$,
\begin{equation}\label{te1}
T^{11}=-\frac{{\cal T}_{1}V}{\sqrt{1-E^{2}}}
\left({\cal F}-\frac{2T'^{2}}{1-E^{2}}{\dot {\cal F}}\right)={\rm constant},
\end{equation}
and the gauge equation (\ref{geq}), $(\Pi)'=0$,
\begin{equation}\label{ge1}
\Pi\equiv \frac{\delta S}{\delta (\partial_{0}A^{1})}=
-\frac{{\cal T}_{1}VE}{\sqrt{1-E^{2}}}
\left({\cal F}-\frac{2T'^{2}}{1-E^{2}}{\dot {\cal F}}\right)={\rm constant}.
\end{equation}
Since the constant pressure $T^{11}$ (\ref{te1}) and the constant
fundamental string charge density $\Pi$ (\ref{ge1})
lead to constant electric field $E=\Pi/T^{11}$, we have a single
nontrivial equation, (\ref{te1}) or (\ref{ge1}),
and the obtained solutions can be
classified by two constant parameters $\Pi$ and $E$.

In the following, it is convenient to introduce a rescaled spatial coordinate
\begin{equation}
\tilde{x}=\sqrt{1-E^{2}}\, x.
\end{equation}
Then, with the static ansatz (\ref{an}), the variable $y$ defined in (\ref{y})
becomes
\begin{equation}\label{cof}
y=\frac{T'^{2}}{1-E^{2}} \equiv \tilde{T}'^{2}.
\end{equation}

With the rescaled coordinate $\tilde x$, the action with constant
electric field takes the form
\begin{eqnarray}
\frac{S}{-\int dt}&=& {\cal T}_{1}\int dx \, e^{-\frac{T^{2}}{4}}
\sqrt{1-E^{2}}\,{\cal F}\left(\frac{T'^{2}}{1-E^{2}}\right)
\label{ya1} \nonumber\\
&=& {\cal T}_{1}\int d\tilde{x} \, e^{-\frac{T^{2}}{4}}
\,{\cal F}(\tilde{T}'^{2}),
\end{eqnarray}
which is the same as that for the pure tachyon case.
It means that there is a one-to-one correspondence between a kink
solution in this $E^{2}<1$ case
and that in the pure tachyon case in section 3.1.
Indeed, the form of the equation with rescaled variables is
the same up to an overall constant
\begin{eqnarray}
-T^{11}=\frac{-\Pi}{E}&=&\frac{{\cal T}_{1}V(T)}{\sqrt{1-E^{2}}}G(y),
\end{eqnarray}
which is rewritten as
\begin{equation}\label{Epa}
{\cal E}_{E}=(1-E^{2})K(y)+U(T),
\end{equation}
with ${\cal E}_{E}=-(1-E^{2})$.
Note that, in this case, ``energy'' ${\cal E}_E$ depends on the electric field.
By varying the value of $E$, we can obtain different types of solutions.

(i) \underline{{\bf Array of kink-antikink}}:
When magnitude of the electric field
is strictly less than the critical value, $E^2 < 1$, the situation is
essentially the same as in section 3. That is, when
$-T^{11}<{\cal T}_{1}/\sqrt{1-E^{2}}$, we have
array of kink-antikinks as the unique regular static soliton
solution of codimension-one, where the tachyon field oscillates between
$\pm {\tilde T}_{{\rm max}}=\pm 2\sqrt{\ln
[{\cal T}_{1}/(-T^{11}\sqrt{1-E^{2}}\,)]}\,$.

It is a simple matter to estimate the domain size ${\tilde \xi}$, i.e.,
the distance between a kink and an adjacent antikink, since the only
difference from the previous section is that the spatial coordinate
is rescaled by the factor $\sqrt{1-E^2}$. For fixed $E$, we find that
the distance $\tilde\xi$ has a maximum
${\tilde \xi}_{\max} =2\pi\sqrt{2\ln 2}/\sqrt{1-E^{2}}$ in the vanishing
${\tilde T}_{\max}$ limit
($-T^{11}\rightarrow {\cal T}_{1}/\sqrt{1-E^{2}}$\,). As $-T^{11}$ decreases,
the size $\tilde\xi$ becomes smaller and eventually goes to zero
in the infinite ${\tilde T}_{\max}$ limit ($T^{11}\rightarrow 0$).
This is the same behavior as in the pure tachyon case.

Note, however, that $E$ can be changed to control the the rescaling factor
$\sqrt{1-E^2}$ in the denominator. If the critical limit $E\rightarrow 1$
is taken first for fixed $T^{11}$, one would find that $\tilde\xi$ diverges,
which will be discussed shortly. This implies that there exists a sort
of limit that both $E\rightarrow 1$ and $T^{11}\rightarrow 0$ are taken
simultaneously in such a way that the distance $\tilde\xi$ remains
finite even when $T^{11}\rightarrow 0$ (or $\Pi\rightarrow 0$ equivalently).
A short analysis shows that it happens when the limits are taken keeping
\begin{equation}
(1-E^2) \ln |T^{11}\sqrt{1-E^2}| = \mbox{finite}.
\end{equation}
In other words, in this case one can obtain array solutions with arbitrary
period even in vanishing $T^{11}$ limit in which the amplitude of tachyon
profile becomes infinite. This is somewhat similar to the case of DBI type
EFT with $1/\cosh$ potential \cite{Kim:2003in, Kim:2003ma}.

{}From 00-component of the energy-momentum tensor (\ref{tmn}), we read
energy density given by a sum of constant piece and localized piece
\begin{eqnarray}\label{ed1}
T_{00}&=&\frac{{\cal T}_{1}V(T)}{\sqrt{1-E^{2}}} G(y) \nonumber\\
&=&-\Pi E +{\cal T}_{1}V(T)\sqrt{1-E^{2}}{\cal F}(y).
\end{eqnarray}
The first constant term stands for the energy density of
fundamental string. The second term is localized at the position of
 (anti)kink and, on integrating over the distance $\tilde\xi$, becomes
\begin{eqnarray}
\int_{-{\tilde \xi}/2}^{{\tilde \xi}/2}dx
          {\cal T}_{1}V(T)\sqrt{1-E^{2}}{\cal F}(y)
&=& \int_{-{\tilde \xi}/2}^{{\tilde \xi}/2}d\tilde x\,
        {\cal T}_{1}V(T){\cal F}(\tilde T'^2) \nonumber \\
&=& 2\pi {\cal T}_{1}   \qquad(\mbox{in large $T_{\max}$ limit}).
\end{eqnarray}
Note that in the first line the factor $\sqrt{1-E^2}$ is naturally
incorporated in $\tilde x$. In the second line we have used
the result of the previous section, since it is the same form as (\ref{ede}).
Therefore we again obtain the decent relation
$ {\cal T}_0 = 2\pi {\cal T}_{1} $,
and the obtained array of kink-antikink is identified as
array of D0${\bar {\rm D}}$0.

\begin{figure}[ht]
\begin{center}
\scalebox{1}[1]{\includegraphics{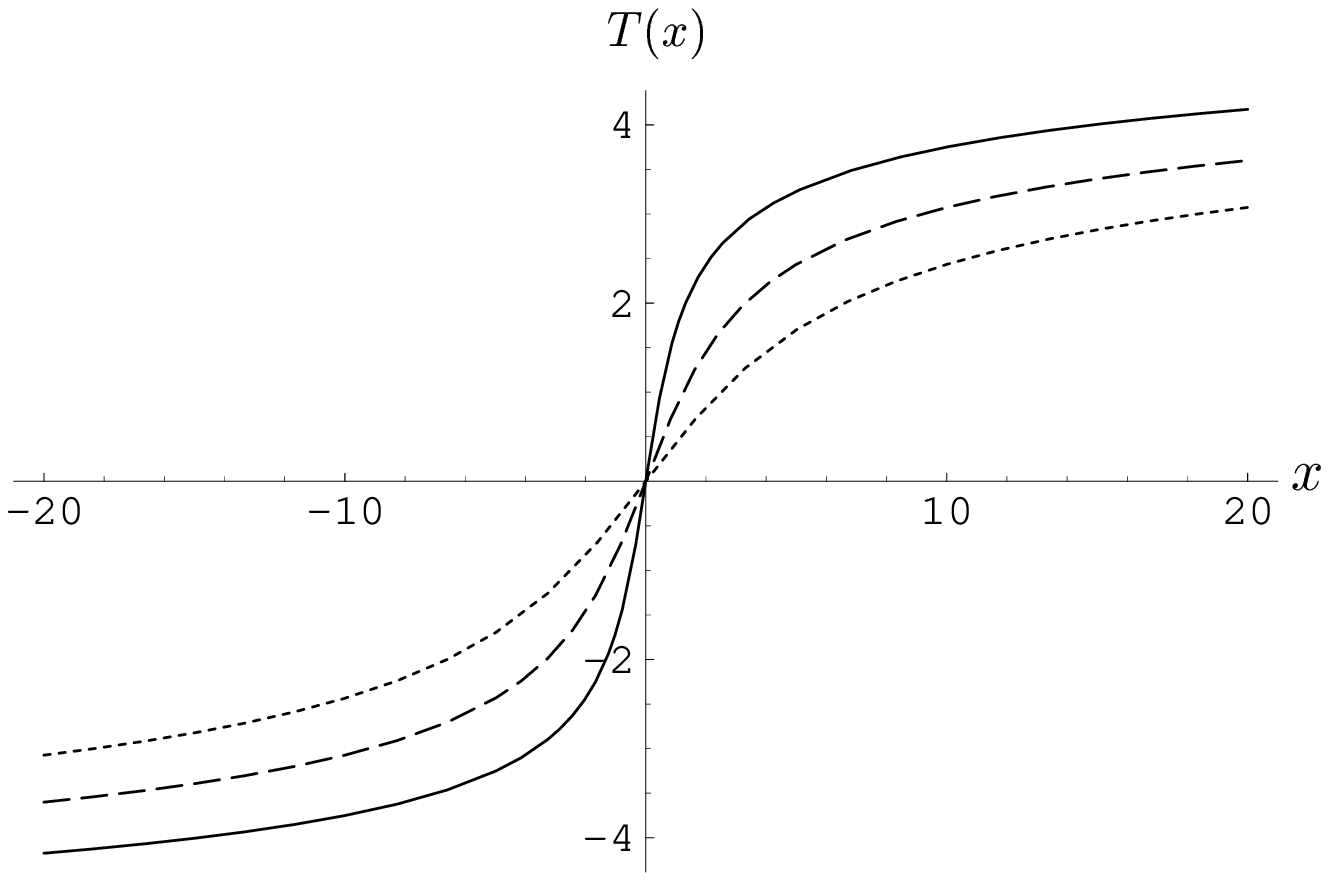}}
\par
\vskip-2.0cm{}
\end{center}
\caption{\small
Profiles of single topological BPS kink for the cases
$\Pi=-1.0\;\mbox{(dotted curve)},\;-0.5\;\mbox{(dashed curve)},\;
-0.2\;\mbox{(solid curve)}$ as the slopes become steep.}
\label{fig4}
\end{figure}

(ii) \underline{\bf Topological BPS kink}:
When the electric field has critical value $E^{2}=1$,
$-\det(\eta+F)=1-E^{2}$ vanishes and $y$ in (\ref{cof})
diverges. Physical quantities, however, have well-defined finite limit and
actually become quite simplified.
To deal with this case properly, we use the following
asymptotic form of ${\cal F}(y)$ for large $y$,
\begin{eqnarray}
{\cal F}({ y})\approx \sqrt{\pi}{ y}^{\frac{1}{2}}
+\frac{\sqrt{\pi}}{8}{ y}^{-\frac{1}{2}}
+{\cal O}({ y}^{-\frac{3}{2}}).
\end{eqnarray}
Then, for example, the action in (\ref{ya1}) becomes
\begin{eqnarray}
\frac{S|_{E^{2}=1}}{-\int dt}&=& \lim_{E^{2}\rightarrow 1}
{\cal T}_{1}\int_{-\infty}^{\infty} dx \, e^{-\frac{T^{2}}{4}}
\sqrt{1-E^{2}}\left[\frac{\sqrt{\pi}|T'|}{\sqrt{1-E^{2}}}
+\frac{\sqrt{\pi (1-E^{2})}}{8|T'|} +{\cal O}((1-E^{2})^{\frac{3}{2}})
\right]
\nonumber\\
&=& \sqrt{\pi}\, {\cal T}_{1}\int_{-\infty}^{\infty}
dx \,e^{-\frac{T^{2}}{4}} |T'| \nonumber\\
&=& \sqrt{2}\, \pi \sqrt{\alpha'}
{\cal T}_{1}={\cal T}_{0}.
\label{de1}
\end{eqnarray}
The equations of motion (\ref{te1}) and (\ref{ge1}) reduce to
\begin{equation}\label{feq}
\frac{-\Pi}{E}=-T^{11}=\frac{\sqrt{\pi}\, {\cal T}_{1}V}{4|T'|}.
\end{equation}
Solving this equation, we obtain an exact solution satisfying
$T'(x=\pm\infty)=0$ in a closed form,
\begin{eqnarray}\label{sk}
T(x)=\pm 2{\rm erfi}^{-1}\left(\frac{{\cal T}_{1}}{4\Pi}x\right),
\end{eqnarray}
where erfi($x$) is the imaginary error function. The profile of the solution
is shown in Fig.~\ref{fig4}.
Since the solution connects monotonically two vacua at positive
and negative infinity, $T(-\infty)=\mp\infty$
and $T(\infty)=\pm\infty$, the obtained solution is interpreted as
single topological tachyon kink (antikink) for the upper (lower) sign.

\begin{figure}[ht]
\begin{center}
\scalebox{1}[1]{\includegraphics{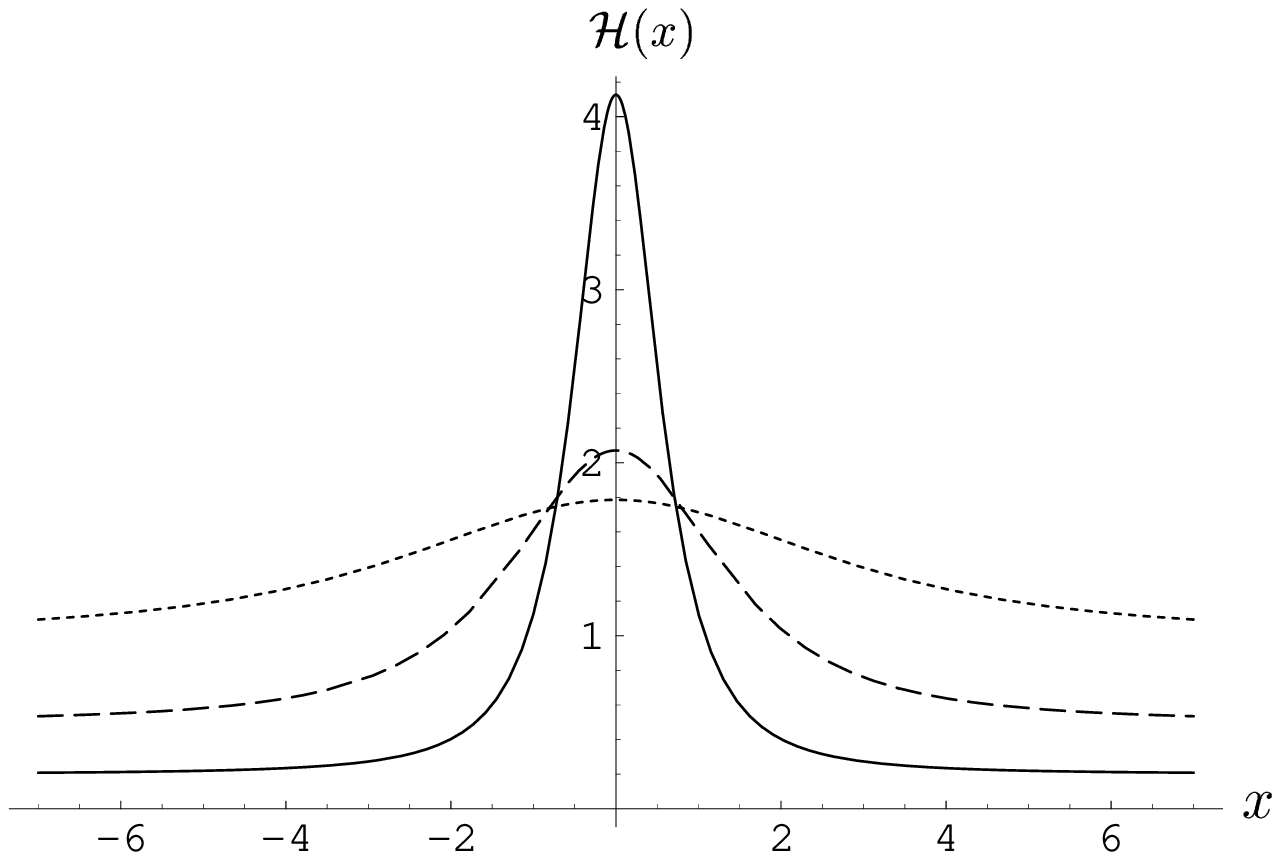}}
\par
\vskip-2.0cm{}
\end{center}
\caption{\small
Energy density profiles of single BPS kink
for the cases $\Pi=-1.0\;\mbox{(dotted curve)},\;-0.5\;\mbox{(dashed curve)},\;
-0.2\;\mbox{(solid curve)}$ as the slopes become steep. It is clear that
their width is decreasing as $T_{\rm max}$ increases.}
\label{fig5}
\end{figure}

The corresponding energy density (\ref{ed1}) is again
written as a sum of a constant part and a localized part
as shown in Fig.~\ref{fig5},
\begin{equation}
T_{00}= - \Pi + \sqrt{\pi}\, {\cal T}_{1}V(T)|T'|.
\label{esu}
\end{equation}
The localized part is the same as that appearing in (\ref{de1}).
As $\Pi$ goes to zero it becomes sharply peaked near $x=0$ and
approaches $2\pi {\cal T}_1\delta(x)$. The energy is then given by a sum of
the D0 charge and the fundamental string charge,
$\Pi$,
\begin{equation}\label{BPS1}
E\equiv \int^{\infty}_{-\infty}dx\, T_{00}
={\cal T}_{0}+Q_{{\rm F1}},
\end{equation}
which is nothing but a BPS sum rule. Note that this result is obtained
regardless of the value of $\Pi$ which controls the thickness of the kink
as seen in the Fig.~\ref{fig5}. The solution can be identified as
a thick single topological BPS tachyon kink discussed
in DBI type EFT \cite{Kim:2003in} and in BCFT~\cite{Sen:2003bc}.
Monotonicity of the tachyon profile for the topological BPS kink (\ref{sk})
shown in Fig.~\ref{fig4} may suggest a tip of hint on the field redefinition
among the tachyon fields in DBI type EFT, BCFT, and BSFT, in which the
profiles of topological BPS kink are given by closed functional form.
As we shall see, the corresponding energy density profiles are slightly
different so that this naive comparison definitely needs further systematic
studies.

\subsection{Arbitrary $p\ge 2$}
The analysis on the unstable D1-brane in the previous section
can be easily generalized to unstable D$p$-branes with $p\ge 2$.
The only difference is that the Bianchi identity,
$\partial_{\mu}F_{\nu\rho}+\partial_{\nu}F_{\rho\mu}+\partial_{\rho}
F_{\mu\nu}=0$ is no longer trivial. However, we show shortly that
all components of the field strength tensor (\ref{fmn})
are just constant, and static kink configurations
are governed by a single first-order differential equation
for the tachyon field,
which is similar to the one for the $p=1$ case. On the other hand,
possible solutions are richer than $p=1$ case;
there exist three other types of solutions in addition to the array of
kink-antikink and the thick single topological BPS kink
in the previous section.

Under the assumption of homogeneous codimension-one objects,
the conservation of energy-momentum (\ref{emc}) becomes $(T^{1\nu})'=0$,
that is, $T^{1\nu}$ are constant.
The expression for $T^{1\nu}$ simplifies to
\begin{equation}\label{c1}
T^{1\nu}=\frac{{\cal T}_{p}V}{\sqrt{\beta}}C^{1\nu}_{{\rm S}}G(y),
\end{equation}
where $\alpha=-C^{11}$ and $\beta=-\det(\eta_{\mu\nu}+F_{\mu\nu})$ with
$C^{\mu\nu}_{{\rm S}}$ denoting symmetric part of the cofactor of
the matrix $(\eta+F)$
and $y=\alpha T'^{2}/\beta$.
The equation for the gauge field (\ref{geq}), $(\Pi^{1\nu})'=0$,
also implies constant-valued $\Pi^{1\nu}$ whose expression is
\begin{equation}\label{c2}
\Pi^{1\nu}=\frac{{\cal T}_{p}V}{\sqrt{\beta}}
C^{1\nu}_{{\rm A}}G(y),
\end{equation}
where $C^{1\nu}_{{\rm A}}$ is the anti-symmetric part of the cofactor.
From these, we conclude that
\begin{equation}
\frac{{\cal T}_{p}V}{\sqrt{\beta}}\label{aaa}
C^{1\nu}G(y)= \mbox{constant}.
\end{equation}

The Bianchi identity imposes $p(p-1)/2$ constraints on the
field strength tensor with $p(p+1)/2$ components. With simple checking,
it is easy to
find that $E^{i}=-F_{0i}$ and $F_{ij}$($i,j\ne 1$) are constants.
In (\ref{c1}) with $\nu=1$, the symmetric part of the cofactor
$C^{11}_{{\rm S}}$ is constant since it is composed of only
$E^{i}$ and $F_{ij}$ ($i,j\ne 1$).
Then (\ref{c1}) tells us that
\begin{equation}\label{c3}
\frac{{\cal T}_{p}V}{\sqrt{\beta}}G(y)=\gamma_p={\rm constant},
\end{equation}
with which (\ref{aaa}) again implies that $C^{1\nu}$ are constant for all
$\nu$'s. When all components of cofactor $C^{1\nu}$ are constant,
$F_{1\nu}$'s are also constant as far as the determinant
$\beta$ does not vanish. The case where the determinant vanishes
can be treated
in a similar fashion as was done in the case
of $p=1$ with $E^{2}=1$.
In summary, we have proved that all components of the field
strength tensor $F_{\mu\nu}$ are constant, and
the only remaining first-order differential equation (\ref{c3})
maybe rewritten as
\begin{equation}\label{c4}
{\cal E}_{p}=K_{p}(y)+U_{p}(T),
\end{equation}
where
\begin{equation}\label{c5}
{\cal E}_{p}=-\frac{\beta}{2\alpha},\qquad
K_{p}(y)=\frac{\beta}{2\alpha}\left[\frac{1}{(G(y))^2}
-1\right],\qquad
U_{p}=-\frac{{\cal T}_{p}^{2}V^{2}}{2\gamma_p^{2}\alpha}.
\end{equation}
This is our primary equation whose solutions we are going to discuss
in the following.
Physical properties of the solutions are characterized by the
following quantities.
Charge density of the fundamental strings along the $x$-direction
is a constant of motion,
\begin{equation}\label{pip}
\Pi^{1}=\frac{{\cal T}_{p}V}{\sqrt{\beta}}
C^{01}_{{\rm A}}G(y)= \gamma_p C^{01}_{{\rm A}}.
\end{equation}
A new aspect of $p\ge 2$ case is that there are $p-1$ spatial directions
orthogonal to $x$. The charge densities of the fundamental string
along these directions are $x$-dependent,
\begin{equation}\label{pia}
\Pi^{a}=\frac{{\cal T}_{p}V}{\sqrt{\beta}} \left[
C^{0a}_{{\rm A}} {\cal F}-2\frac{T'^2}{\beta}\dot{\cal F}
(C^{10}_{{\rm A}}C^{1a}_{{\rm S}}-C^{1a}_{{\rm A}}C^{10}_{{\rm S}})\right],
\qquad (a=2,3,...,p).
\end{equation}
Using (\ref{pip}), these can be written as a sum of a constant part
proportional to $\Pi^1$ and an $x$-dependent part as in (\ref{ed1}),
\begin{equation} \label{pia2}
\Pi^a = \frac1{\alpha C_{\rm A}^{01}}(C_{\rm A}^{10}
C_{\rm S}^{1a} - C_{\rm A}^{1a}C_{\rm S}^{10})\Pi^1
+ (\alpha C_{\rm A}^{0a}-C_{{\rm A}}^{10}C_{\rm S}^{1a}
+C_{{\rm A}}^{1a}C_{\rm S}^{10} )
\frac{{\cal T}_pV}{\alpha\sqrt\beta}{\cal F}.
\end{equation}
Similarly, the expression for the energy density is given by
\begin{eqnarray}\label{ep}
{\cal H}=T_{00} &=& \frac{{\cal T}_{p}V}{\sqrt{\beta}} \left(
C^{00}_{{\rm S}} {\cal F}+2\frac{T'^2}{\beta}\dot{\cal F}
C^{01}C^{10}\right) \nonumber \\
&=& \frac{C^{01}C^{10}}{\alpha C_{{\rm A}}^{10}} \Pi^1
+ \left(C^{00}+\frac1\alpha C^{01}C^{10} \right)
\frac{{\cal T}_pV}{\sqrt\beta}{\cal F}.
\end{eqnarray}
Comparing (\ref{pia2}) and (\ref{ep}), we notice that the localized pieces
share the same functional shape except for overall coefficients. Therefore,
it is enough to look into either the Hamiltonian density (\ref{ep})
or the charge density (\ref{pia2}) when properties of the localized
pieces are discussed.

Now we find the solution.
Since BSFT action (\ref{yac}) is valid only
for nonnegative
$\beta=-\det(\eta_{\mu\nu}+F_{\mu\nu})$,
there are five different cases to consider,
depending on the sign of $\alpha$ and the magnitude of $\beta$.

(i) \underline{{\bf Array of kink-antikink for $\alpha>0$ and $\beta>0$}}:
When both $\alpha$ and $\beta$ are positive, the mechanical energy
${\cal E}_{p}$ and the potential energy $U_{p}(T)$ in (\ref{c5}) are
negative, and our equation (\ref{c4}) is identical to the equation
(\ref{ene}) for the pure tachyon case in the previous section
up to a trivial rescaling of the spatial coordinate,
$\tilde x=\sqrt{(\beta/\alpha)}\,x$ as done in $p=1$ case. We then obtain
kink-antikink array solutions. The second term of the energy
density (\ref{ep}) is localized at the sites of kinks and antikinks.
Note that the charge densities $\Pi^a$ in (\ref{pia2}), which orthogonal
to $x$-direction, also have peak values at (anti)kink sites.
This may be interpreted as fundamental strings confined on
D$(p-1)\,\bar{\rm D}(p-1)$ branes.

\begin{figure}[ht]
\begin{center}
\vspace{10mm}
\scalebox{1}[1]{\includegraphics{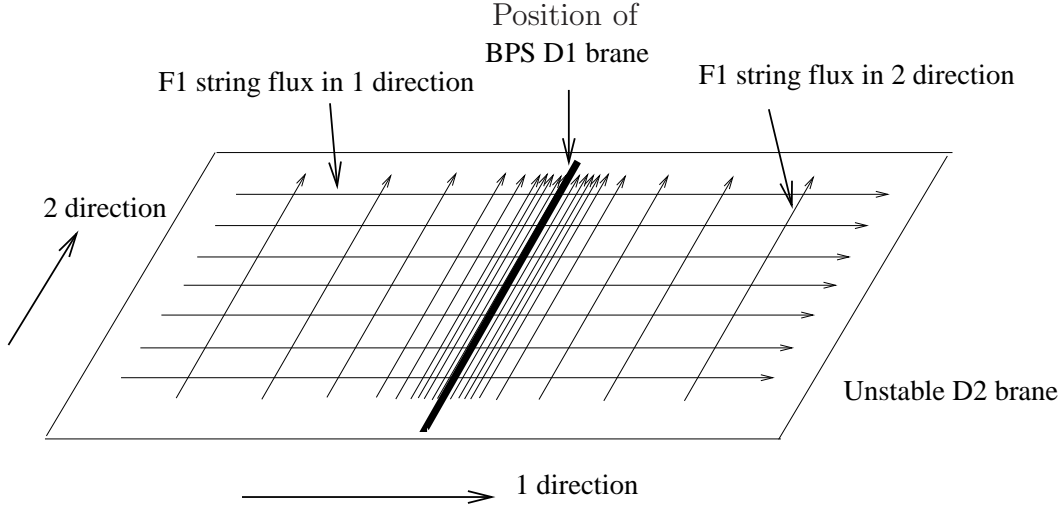}}
\par
\vskip-2.0cm{}
\end{center}
\caption{\small A schematic figure of confined fundamental strings along
stable D1-brane from an unstable D2-brane.}
\label{fig6}
\end{figure}

(ii) \underline{{\bf Topological BPS kink for $\alpha>0$ and $\beta=0$}}:
Similarly, $\beta = 0$ corresponds to
the critical electric field $E^2=1$ in the $p=1$ case and we obtain a single
BPS kink solution in closed form (\ref{sk}) with a suitable change
of the parameters.
Again the only difference is the localization of fundamental strings
on the codimension-one D-brane so that the obtained topological BPS
kink is D$(p-1)$F1 composite as shown in Fig.~\ref{fig6}.
We omit the details.

\begin{figure}[ht]
\begin{center}
\scalebox{0.6}[0.6]{\includegraphics{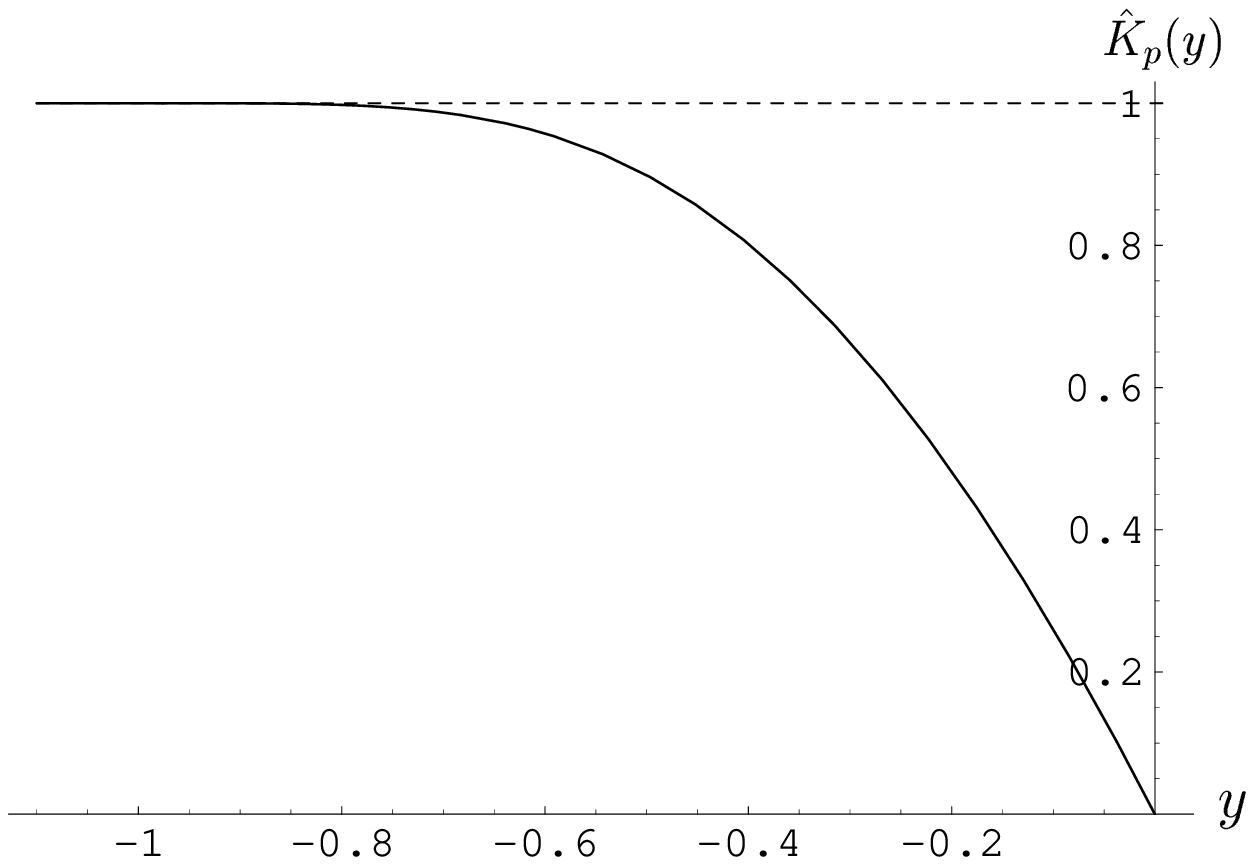}
\includegraphics{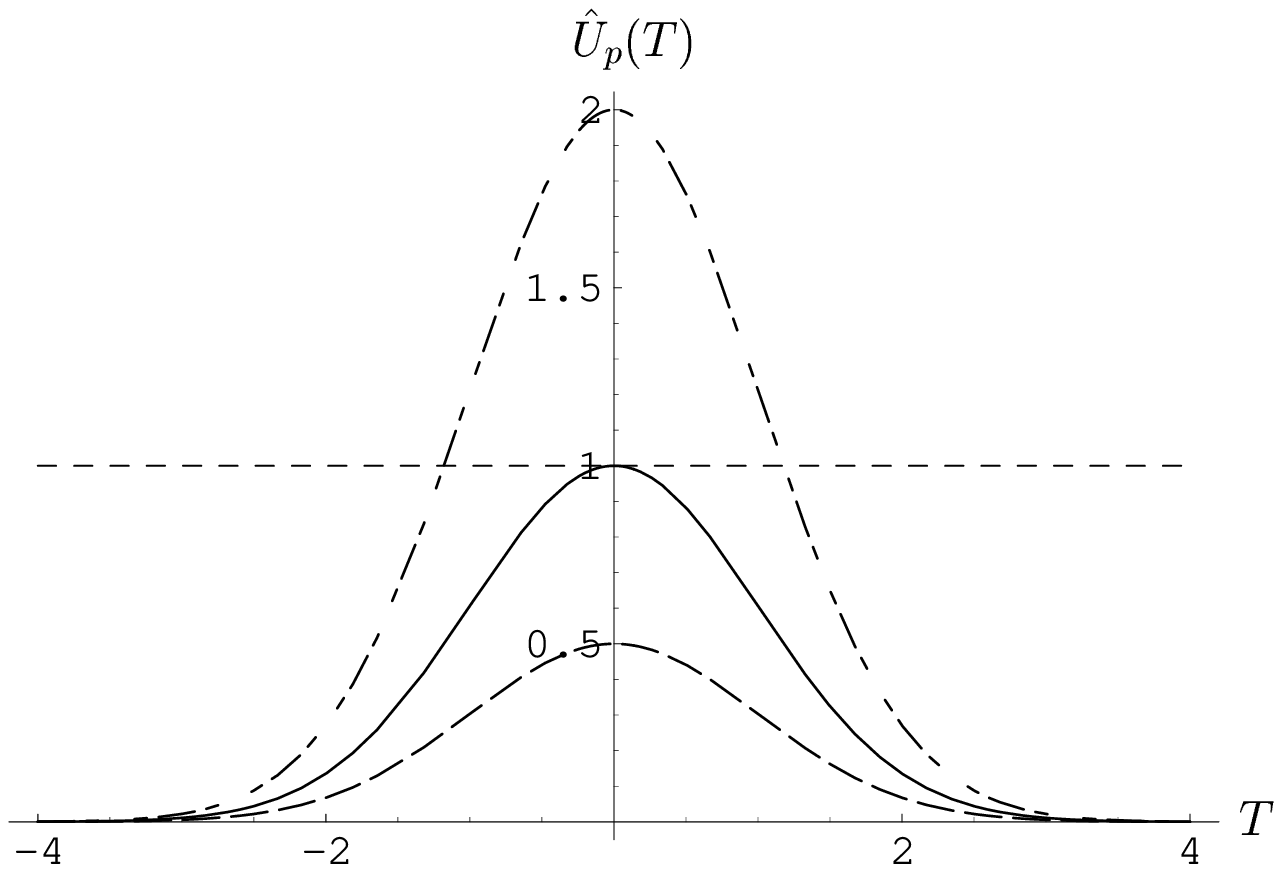}}
\par
\vskip-2.0cm{}
\end{center}
\caption{\small The graphs of ${\hat K}_{p}(y)$
and ${\hat U}_{p}(T)$.  For ${\hat U}_{p}(T)$, three cases of
${\cal T}_p /(\gamma_{p}\sqrt{\beta})$ is larger than (dotted-dashed curve,
$1/ \sqrt{2}$), equal to (solid curve, 1), and smaller than
(dashed curve, $\sqrt{2}$) unity
from top to bottom are shown. ${\hat {\cal E}}_{p}=1$ is given by
a straight dotted lines in both figures.}
\label{fig7}
\end{figure}

Although both the array of kink and antikink and the single topological BPS
kink exist on an unstable D$p$-brane for any $p$,
the next three cases that we are going to discuss
are supported only for $p\ge 2$ and negative $\alpha$.
For these cases, it is convenient to multiply $-\alpha/\beta$ to both sides
of our primary equation (\ref{c4}), so that the resultant equation coincides
formally with that of the homogeneous rolling tachyons~\cite{IKK}
\begin{equation}\label{c6}
{\hat {\cal E}}_{p}={\hat K}_{p}(y)+{\hat U}_{p}(T),
\end{equation}
with,
\begin{equation}\label{c7}
{\hat {\cal E}}_{p}=1,\qquad
1\ge {\hat K}_{p}(y)=1-\frac{1}{G(y)^2}\ge 0,\qquad
{\hat U}_{p}=\frac{{\cal T}_{p}^{2}V(T)^{2}}{\gamma_p^{2}\beta}\ge 0,
\end{equation}
for negative $y={\alpha\over \beta}T'^2$. We draw ${\hat K}_{p}$ and
${\hat U}_{p}$ in Fig.~\ref{fig7}.
We use $\beta\ge 0$ for soliton solutions due to
max$({\hat K}_{p})=1$ which requires ${\hat {\cal E}}_{p}-
{\hat K}_{p}(y)={\hat U}_{p}(T)\ge 0$.
Since ${\hat {\cal E}}_{p}$ is fixed to be unity and $d{\hat U}_{p}/dT$
vanishes only at $T=0,\pm\infty$, the domain $-1\le y\le 0$ is
of our interest for
${\hat K}_{p}(y)$ which is monotonic decreasing from
${\hat K}_{p}(-1)=1$ to ${\hat K}_{p}(0)=0$ (see Fig.~\ref{fig7}).

When $\beta=0$, there exist only trivial vacuum solutions at $T=\pm\infty$
and thereby we consider only positive $\beta$ for nontrivial kinks.
${\hat {\cal E}}_{p}=1$ provides three situations that the top of
${\hat U}_{p}$ is higher than ${\hat {\cal E}}_{p}$
(the dotted-dashed line in Fig.~\ref{fig7},
$0<\beta<({\cal T}_{p}/\gamma_p)^{2}$),
equal to ${\hat {\cal E}}_{p}$
(the solid line in Fig.~\ref{fig7},
$\beta<({\cal T}_{p}/\gamma_p)^{2}=0$),
and lower than ${\hat {\cal E}}_{p}$
(the dashed line in Fig.~\ref{fig7},
$\beta<({\cal T}_{p}/\gamma_p)^{2}=0$). We are discussing those three
cases below.

\begin{figure}[ht]
\begin{center}
\scalebox{0.6}[0.6]{\includegraphics{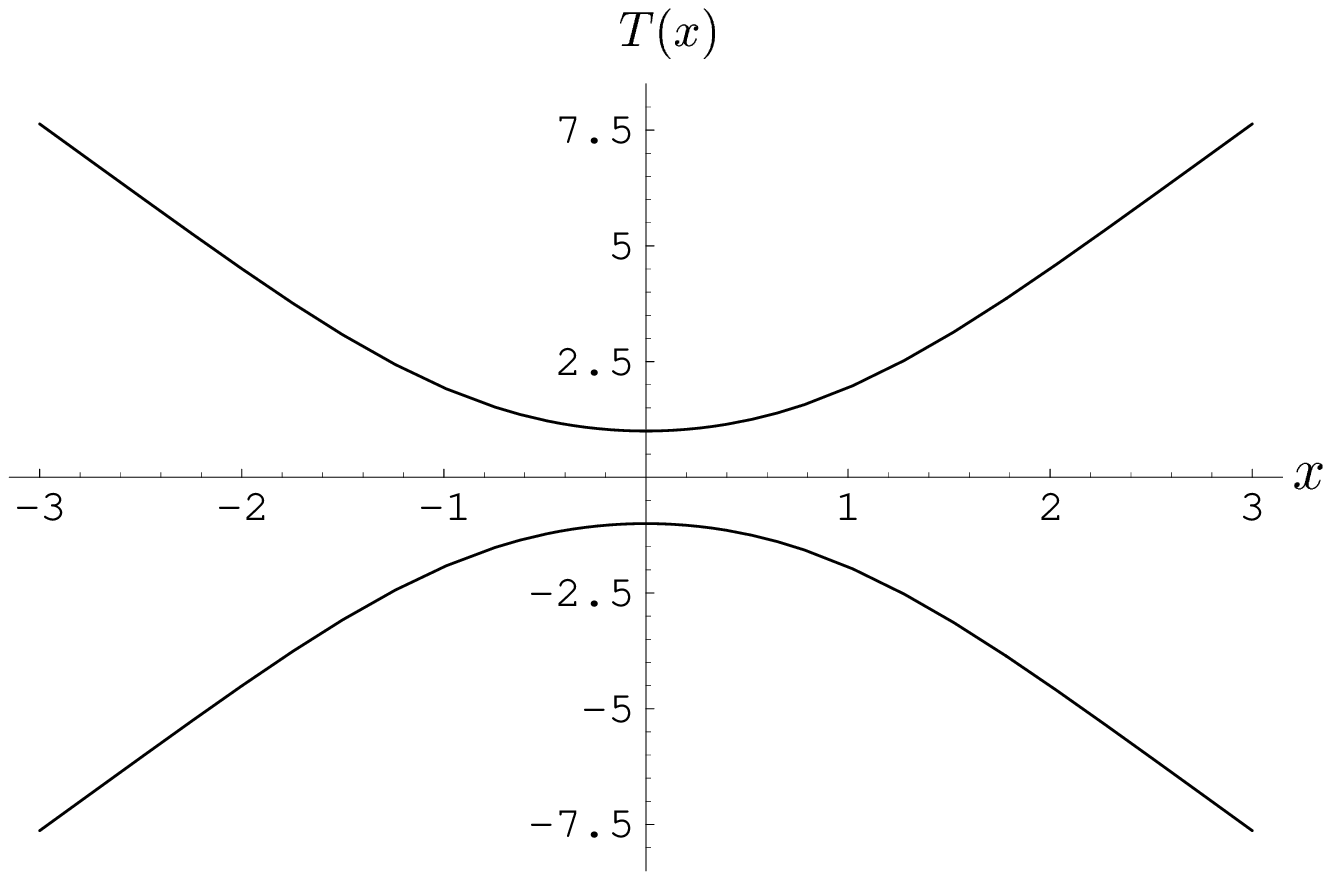}
\includegraphics{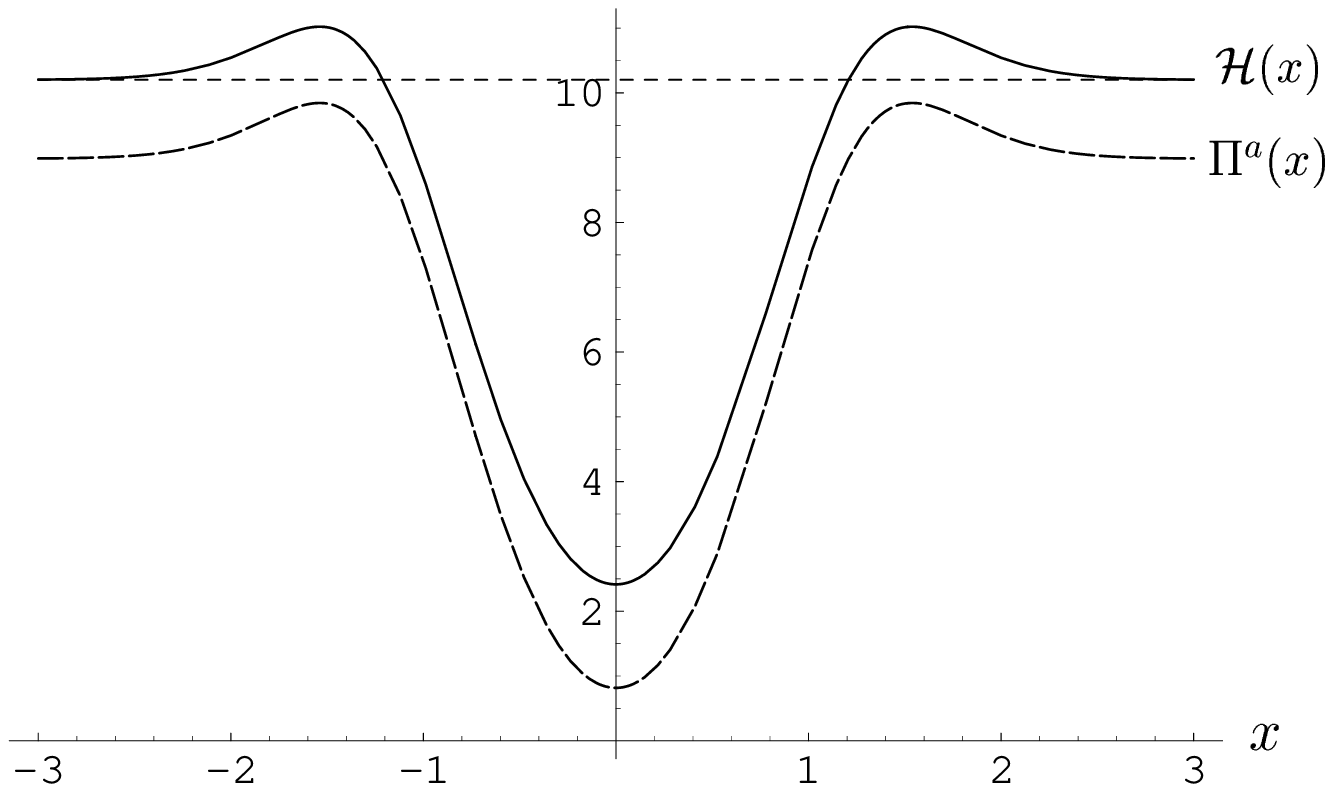}}
\par
\vskip-2.0cm{}
\end{center}
\caption{
{\small The graphs of bounce for ${\cal T}_{p}^{2}/(\gamma_p^{2}
\beta)=1.65>1$. The tachyon field $T(x)$ is given in the left figure.
The energy density ${\cal H}(x)$ (solid line) and the charge density
of fundamental strings along the transverse directions $\Pi^{a}$
(the dashed line) are given in the right figure.}}
\label{fig8}
\end{figure}

(iii) \underline{{\bf Bounce for $\alpha<0$ and
$0<\beta<({\cal T}_{p}/\gamma_p)^{2}$}}:
As shown by the solid line in Fig.~\ref{fig8},
we let the bounce solution have its minimum value at
$x=0$; $T(x=0)=T_{{\rm bounce}}
=\sqrt{2\ln\left(\frac{{\cal T}_p^2}{\beta\gamma_p^2}\right)}$
with $T'(x=0)=0$ (see Fig.~\ref{fig8}).
Expanding the solution near the turning point, we have
\begin{equation}\label{tr}
T(x)\approx T_{{\rm bounce}}+ a_{{\rm b}}  x^2 + {\cal O}(x^4),
\end{equation}
where $a_{{\rm b}} = \frac{\beta T_{{\rm bounce}}}{
16\ln 2\, (-\alpha)}$.
Inserting this into the energy density (\ref{ep}), we get the profile
of the energy density near $x=0$,
\begin{equation}\label{bed}
{\cal H}=T_{00}\approx
\frac{{\cal T}_p\,e^{-\frac{T^2_{{\rm bounce}}}{4}}}{\sqrt{\beta}}
\left\{C^{00}_{{\rm S}}
+ 16\ln 2 \left[ C^{01}C^{10}-(-\alpha) C^{00}_{{\rm S}}
\right]
\frac{a_{{\rm b}}^2}{\beta}\, x^2+  {\cal O}(x^4)\right\}.
\end{equation}
In addition to the positive constant leading term, the coefficient of
subleading term is also positive which is
consistent with numerical result in Fig.~\ref{fig8}.
To see why this is true, consider an exemplar case of $p=2$,
where we have (we let $x=x^1$)
\begin{eqnarray}
\alpha=-C^{11}=1-E_2^2 <0,\qquad \beta=-{\rm det}(\eta+F)
=1+B^2-E_1^2-E_2^2 >0,\nonumber\\
C^{00}=1+B^2,\qquad C^{01}C^{10}=E_2^2 B^2 -E_1^2,
\end{eqnarray}
so that
\begin{equation}
C^{01}C^{10}-(-\alpha)C^{00}=\beta >0.\label{ineq}
\end{equation}
Note also that $C^{01}C^{10}$ is always positive.

Near the infinity $x\rightarrow \pm \infty$,
using the fact that $G( y)$ blows up at $ y\to -1$ as
$G( y)\to {1\over( y+1)^2}-{3\over 2}{1\over (y+1)}$,
the tachyon field behaves as (we only focus on the one side
where $x\to \infty$ or $T\to \infty$ for the other side is symmetric.)
\begin{equation}\label{xin}
T(x)\approx \sqrt{\frac{\beta}{(-\alpha)}}\, \left[(x-x_0)
+{2(-\alpha)\sqrt{{\cal T}_p}\over
\sqrt{\gamma_p}\beta^{5\over 4} }{1\over (x-x_0)}
\,e^{-\frac{\beta}{(-8\alpha)}\, (x-x_0)^2}\right],
\end{equation}
where $x_0$ is determined only by the boundary condition at $x=-\infty$.
Substituting the tachyon profile (\ref{xin}) into the energy density
(\ref{ep}), we obtain the positive constant leading term and
the positive but exponentially-decreasing subleading term as
\begin{eqnarray}\label{h}
{\cal H}&\approx& \frac{\gamma_p}{(-\alpha)}C^{01}C^{10}
+{\sqrt{\gamma_p {\cal T}_p} \over 2(-\alpha)\beta^{1\over 4}}
\left[C^{01}C^{10}-(-\alpha)C^{00}\right]
e^{-{\beta \over 8(-\alpha)}(x-x_0)^2}.
\end{eqnarray}
The obtained asymptotic profile (\ref{h}) is consistent
with numerical result in Fig.~\ref{fig8}.

In the above, if we subtract the asymptotic constant piece,
which is due to condensation
of the fundamental strings on the D$p$-brane represented by the
constant electromagnetic field strength, we can interpret
the resulting energy profile
as the bound energy of a composite of the tachyon bounce
and confined fundamental
strings.
By studying numeric solutions, Fig.~\ref{fig8},
it is interesting to note that the energy and $\Pi^a$ profiles
(after subtracting their asymptotic values at infinity)
have a positive bump at some finite distance from the center.
We see that there exists a position $x_{\ast}$ such that
${\cal H}(x_{\ast})={\cal H}(\pm\infty)$ and
$\Pi^a(x_{\ast})=\Pi^a(\infty)$, although it is not possible to
express it analytically. From the Fig.~\ref{fig8}, we read that a localized
bounce with negative energy is formed by repelling the condensed DBI
electromagnetic field.
Therefore, this composite may be interpreted as a negative energy
{\it brane} of codimension-one, generated through {\it de-condensing} the
background fundamental strings with positive constant energy density.

Note that the obtained asymptotic behavior (\ref{xin}) can also be
applied to other soliton solutions with $\beta>0$ and $\alpha<0$
given below. However they are different from those from DBI type
EFT~\cite{Kim:2003in,Kim:2003ma}.
Since the profiles of energy-momentum tensors in BCFT
and DBI type EFT are qualitatively the same,
we expect that if these soliton configurations are reproduced
in the context of BCFT, their asymptotic behaviors would probably
be slightly different
from what we achieved here in BSFT.

\begin{figure}[ht]
\begin{center}
\scalebox{0.6}[0.6]{\includegraphics{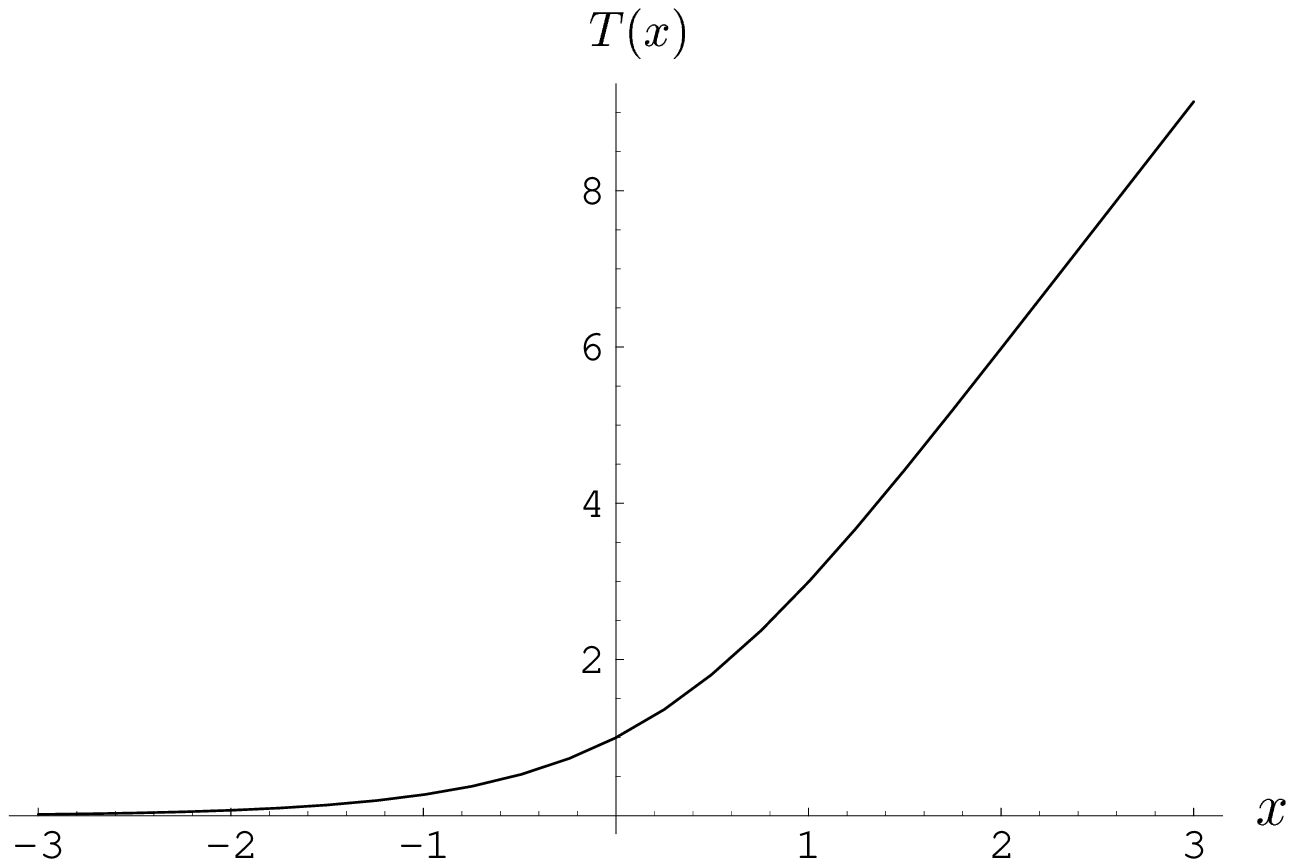}
\includegraphics{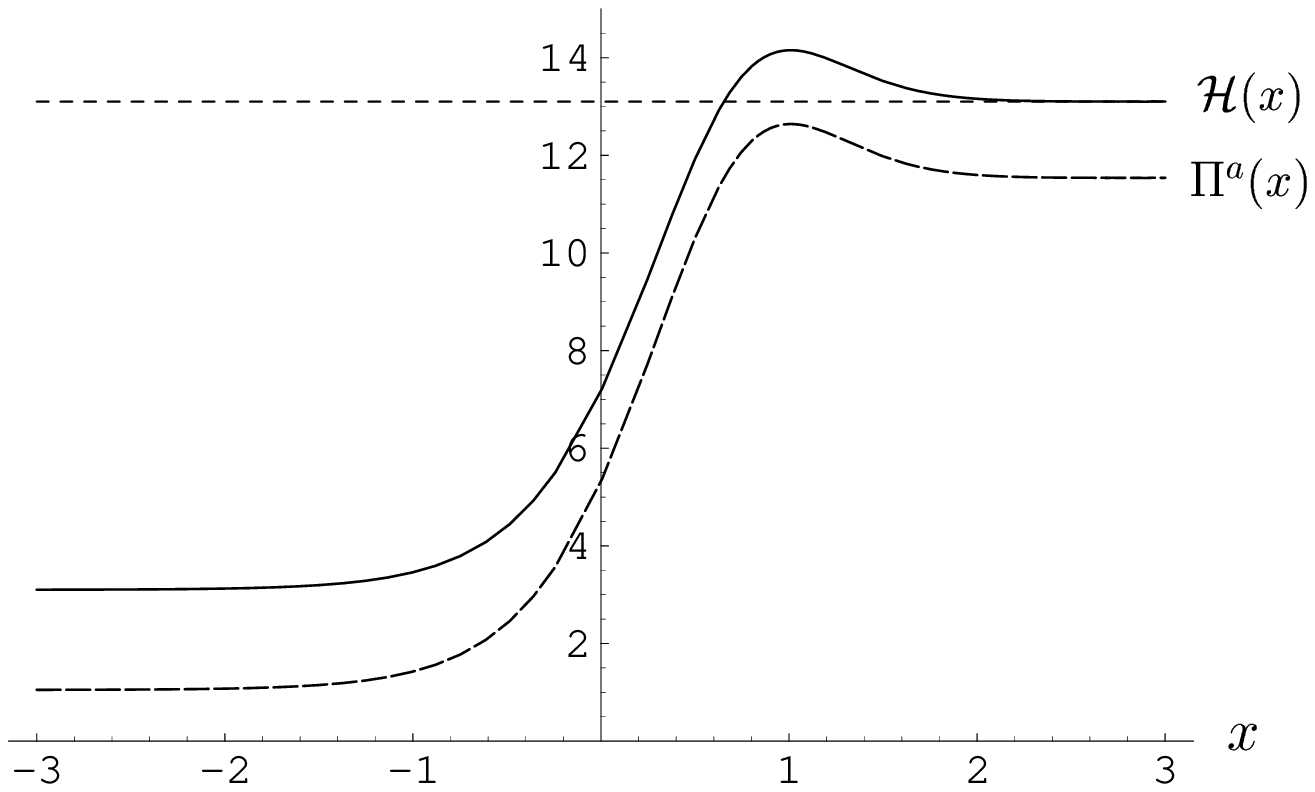}}
\par
\vskip-2.0cm{}
\end{center}
\caption{
{\small The graphs of half kink for ${\cal T}_{p}^{2}/(\gamma_p^{2}\beta)=1$.
The tachyon field $T(x)$ is given in the left figure.
The energy density ${\cal H}(x)$ (solid line) and the charge density
of fundamental strings along the transverse directions $\Pi^{a}$
(dashed line) are given in the right figure.}}
\label{fig9}
\end{figure}

(iv) \underline{{\bf Half-kink for $\alpha<0$ and
$\beta=({\cal T}_{p}/\gamma_p)^{2}$}}:
When $\beta=({\cal T}_{p}/\gamma_p)^{2}$, we
have a half-kink solution satisfying $T(x=-\infty)=0$ and
$T(x=\infty)=\pm\infty$ (see Fig.~\ref{fig9}).
The leading behavior of the tachyon around $x=-\infty$ is exponential
as expected,
\begin{eqnarray}
T(x)\approx e^{a_{{\rm h}} x}+ \cdots,
\end{eqnarray}
and then behavior of the physical quantity (\ref{pia}) (or (\ref{ep})) around
$x=-\infty$ is easily shown to be
\begin{eqnarray}\label{hm}
{\cal H}\approx \frac{{\cal T}_p}{\sqrt{\beta}}
\left[ C^{00}_{{\rm S}}+ \frac{1}{2(-\alpha)} \left(
C^{01}C^{10}-(-\alpha) C^{00}_{{\rm S}}\right)e^{2a_{{\rm h}} x}\right]
+ \cdots,
\end{eqnarray}
where $a_{{\rm h}} = \sqrt{\frac{\beta}{8\ln 2\,(-\alpha)}}$.
In (\ref{hm}), the leading term is constant and
the coefficient of the subleading term for the energy density is
positive as before.
In the exemplar case of $p=2$, we have by (\ref{ineq}),(note
that ${{\cal T}_p \over \sqrt{\beta}}=\gamma_p$ in our case)
\begin{equation}
{\cal H}(-\infty)=\left({\cal T}_p \over \sqrt{\beta}\right)
C_{{\rm S}}^{00}=\gamma_p C^{00}_{{\rm S}}< {\gamma_p \over (-\alpha)}
C^{01}C^{10}={\cal H}(\infty),
\end{equation}
and we again find that there is a point $x_{\ast}$ such that
${\cal H}(x_{\ast})={\cal H}(\infty)$ and $\Pi^{a}(x_{\ast})=
\Pi^{a}(\infty)$.
Also, $C^{02}_{{\rm A}}=-E_2$, and $C^{10}_{{\rm A}} C^{12}_{{\rm S}}
-C^{12}_{{\rm A}} C^{10}_{{\rm S}}=E_2(E_1^2+B^2)$,
so that we have
\begin{equation}
\Pi^2(-\infty)=\left({\cal T}_p \over \sqrt{\beta}\right)
C^{02}_{{\rm A}}=\gamma_p(-E_2)
,\quad \Pi^2(\infty)={\gamma_p \over (-\alpha)}(-E_2)(E_1^2+B^2),
\end{equation}
and they have the same direction.

The half kink configuration connects smoothly
the false symmetric vacuum at $T=0$
and the true broken vacua at $T=\pm\infty$ so that this static object forms
a boundary of two phases. Therefore, it can be interpreted as a
{\it half brane} ($\frac{1}{2}$D($p-1$)-brane)
similar to bubble wall at a given time.
The profile of $\Pi^{a}(x)$ in Fig.~\ref{fig9} suggests that the fundamental
strings that are condensed in the true and the false vacuum align
in the same direction.
In the case of DBI type EFT and NCFT,
after subtracting the vacuum energy from the
unstable vacuum, the total energy of the half kink solution vanishes
exactly~\cite{Banerjee:2004cw}. Though we need careful analysis to answer
this in BSFT, it is at least the configuration of almost zero energy.
Usually a bubble wall produced with radius larger than critical value
starts to expand by consuming vacuum energy of the false vacuum.
Therefore, it may be intriguing to ask such dynamical questions for this
half brane.

\begin{figure}[ht]
\begin{center}
\scalebox{0.6}[0.6]{\includegraphics{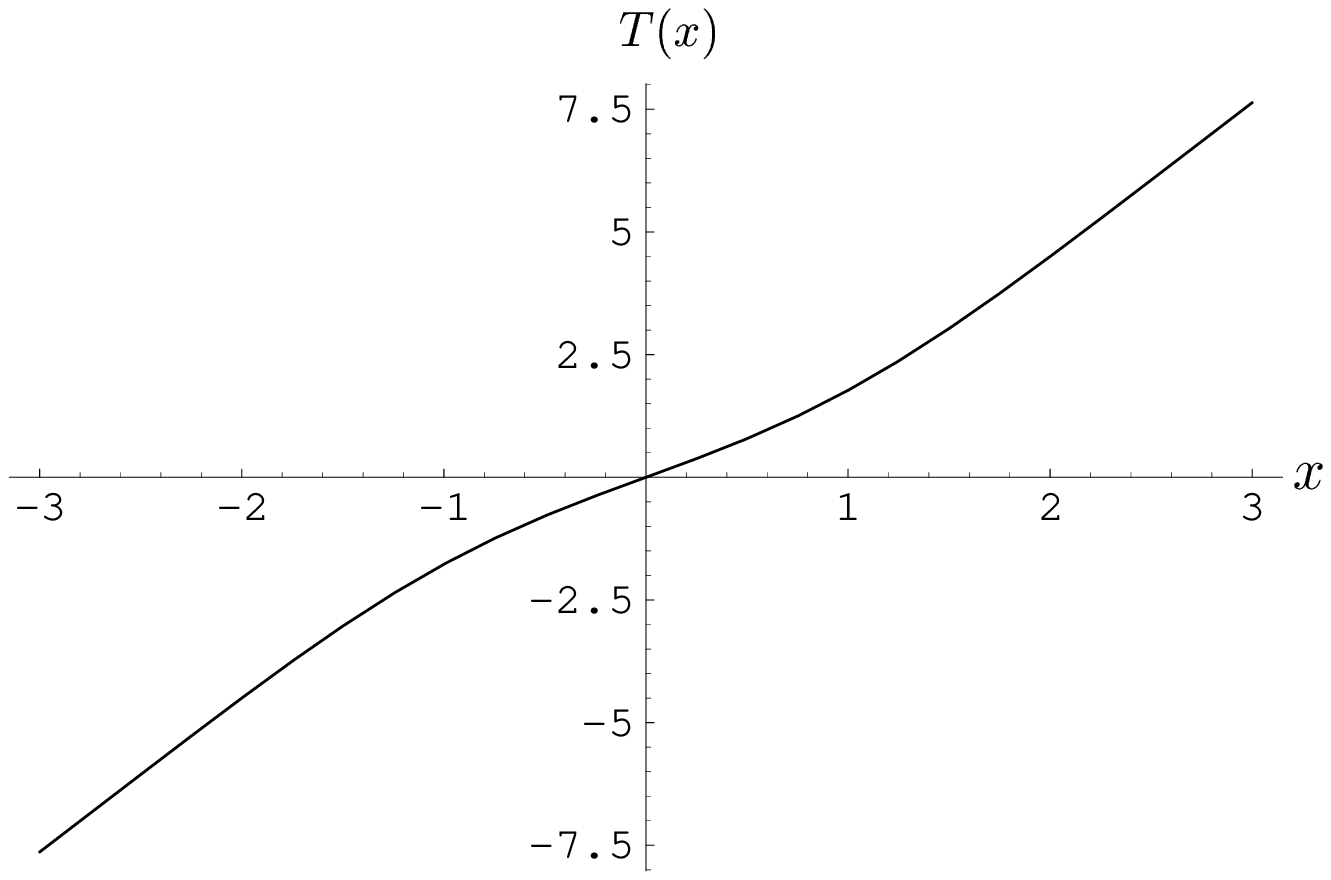}
\includegraphics{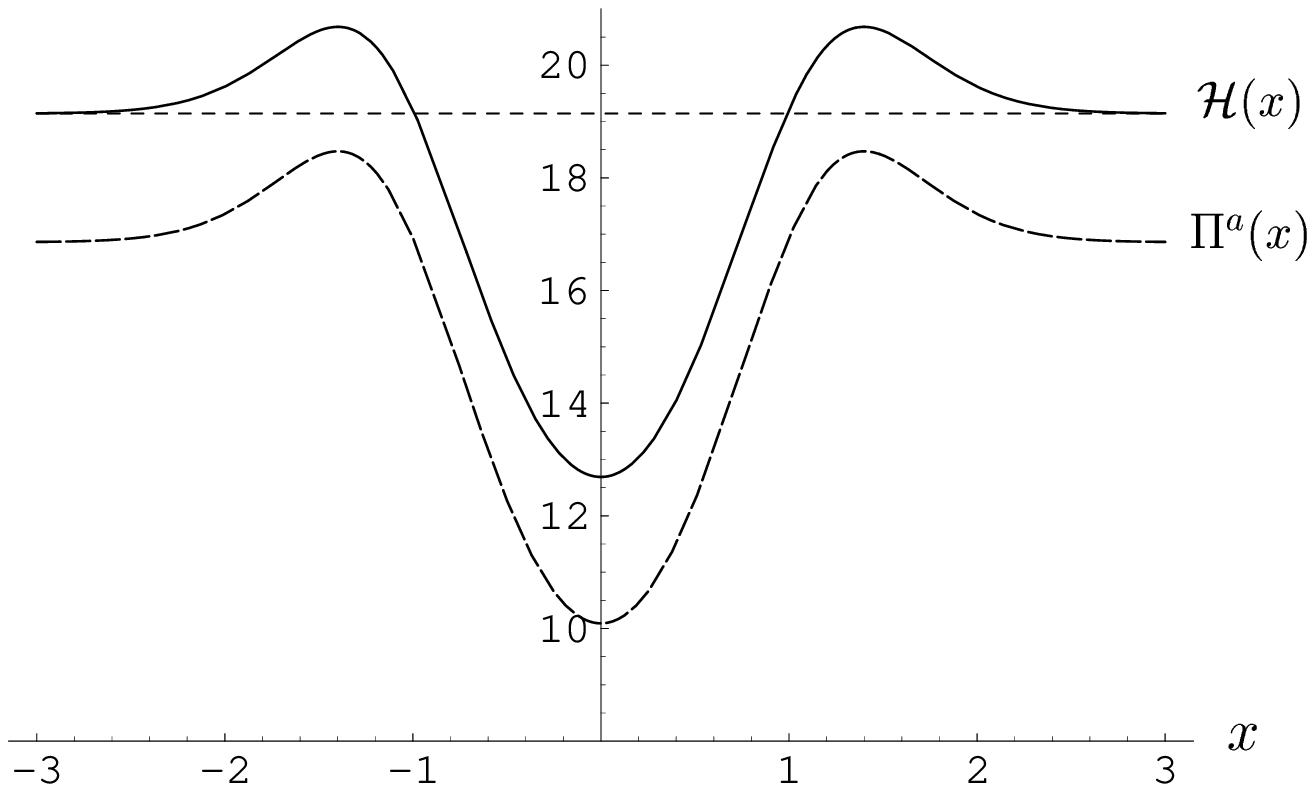}}
\par
\vskip-2.0cm{}
\end{center}
\caption{
{\small The graphs of nonBPS topological kink for
${\cal T}_{p}^{2}/(\gamma_p^{2}\beta)=0.47<1$. The tachyon field $T(x)$ is
given in the left figure.
The energy density ${\cal H}(x)$ (solid line) and the charge density
of fundamental strings along the transverse directions $\Pi^{a}$
(dashed line) are given in the right figure.}}
\label{fig10}
\end{figure}

(v) \underline{{\bf Topological nonBPS kink for $\alpha<0$ and
$\beta>({\cal T}_{p}/\gamma_p)^{2}$}}:
This case describes topological nonBPS kinks connecting both vacua
$T\to \pm\infty$ as shown in Fig.~\ref{fig10}.

If we restrict our analysis to cases where $\beta$ is very
close to $({\cal T}_{p}/\gamma_p)^{2}$
so that $T'(x=0) \ll 1$ when the
tachyon is near the top of the tachyon potential at $x=0$,
we can approximate ${1\over G(y)}\sim 1+(2\ln 2) y$ for $y\ll 1$, and
\begin{eqnarray}
T(x)&\approx& a_{{\rm t}} x + {1\over 3} b_{\rm t}x^3+{\cal O}(x^5),
\end{eqnarray}
where
\begin{eqnarray}
a_{{\rm t}} &=&
\sqrt{\frac{\beta}{2\ln 2(-\alpha)}\left(1-\frac{{\cal T}_p}{\sqrt{\beta}
\gamma_p}\right)}\, ,\nonumber\\
b_{\rm t}&=& a_{\rm t} {{\cal T}_p
\sqrt{\beta} \over (16\ln2)(-\alpha)\gamma}.
\end{eqnarray}
Substituting this into the energy density (\ref{ep}), we obtain near $x=0$,
\begin{eqnarray}
{\cal H}&\approx& \frac{{\cal T}_p}{\sqrt{\beta}}\left[
C^{00}_{{\rm S}}\left( \frac{{\cal T}_p}{\sqrt{\beta}
\gamma_p}\right)
+ \frac{2}{(-\alpha)}\left(1-\frac{{\cal T}_p}{\sqrt{\beta}
\gamma_p}\right)C^{01}C^{10}\right]
\\
&&\hspace{-10mm}-\frac{{\cal T}_p}{\sqrt{\beta}}
{\beta \over (4\ln 2)(-\alpha)}\left(1-\frac{{\cal T}_p}{\sqrt{\beta}
\gamma_p}\right)\left[C_S^{00}\frac{{\cal T}_p}{\sqrt{\beta}
\gamma_p}+{C^{01}C^{10} \over (-\alpha)}\left(1-\frac{2{\cal T}_p}{\sqrt{\beta}
\gamma_p}\right)\right] x^2 .
\label{toh}
\end{eqnarray}
Similar to the previous solutions, the leading term in
(\ref{toh}) is constant
and the coefficient of the second term is positive.
Though the profile of the tachyon is different from that of the bounce
in Fig.~\ref{fig8}, the energy density and the string charge density
of the nonBPS topological kink resemble those of the bounce.
Therefore, the same arguments in the case of bounce can be repeated and
we omit those.
Due to their boundary conditions, the bounce may be an unstable object but
the nonBPS topological kink is a stable soliton.
At any rate the obtained nonBPS topological kink can be
interpreted as a {\it brane} with finite negative tension where the
condensed fundamental strings are repelled.

There have been found similar nonBPS topological kink solutions in
DBI type EFT for both $\alpha>0$ and $\beta<0$.
Since the tachyon action
of our interest (\ref{yac}) is not valid in the range of $\beta<0$,
we only find the solutions for $\beta>0$.

\setcounter{equation}{0}
\section{Conclusion and Outlook}

We studied effective theory of a real tachyon and abelian gauge field
in the framework of BSFT, describing a system of an unstable D$p$-brane with
fundamental strings in superstring theory.
Static kink configurations and their interpretation as codimension-one
branes were of our interest.

When  gauge field is absent, there exists a unique solution of the
array of kink-antikink and it is identified as an array of
D($p-1$)${\bar {\rm D}}(p-1)$. This uniqueness coincides with
the results of BCFT~\cite{Sen:2003bc} and DBI type
EFT~\cite{Lambert:2003zr,Kim:2003in}.
In the presence of the gauge field, we found rich spectrum of static
solitons including array of kink-antikink, topological BPS kink,
bounce, half-kink, and nonBPS topological kink.
It is also consistent with the result of DBI type
EFT~\cite{Kim:2003in,Kim:2003ma} and NCFT~\cite{Banerjee:2004cw}.
For the topological BPS kink, we find a closed functional form of
the tachyon field profile. Since we also have exact solution of
this BPS kink in DBI type EFT~\cite{Kim:2003in} and BCFT~\cite{Sen:2003bc},
we may read a tip of
field redefinition between the tachyon field in BSFT and that in BCFT.

In the context of superstring theory, interpretation of some of the above
objects is clear. To be specific, the array of kink-antikink should
be an array of
D$(p-1){\bar {\rm D}}(p-1)$
or that of D$(p-1)$F1${\bar {\rm D}}(p-1)$F1 where some fundamental strings
are confined on D-brane. The topological BPS kink is a
single BPS D$(p-1)$-brane or D$(p-1)$F1 composite out of unstable D$p$-brane.
Our analysis also suggests
an interpretation for other three solitonic configurations.
The bounce connecting
the same unstable vacuum may be an
unstable brane of negative tension formed by repelling the condensed
background fundamental strings. The nonBPS topological kink may  be
a stable brane of negative tension by decondensing fundamental strings on it.
The half-kink must be an intriguing object, which connects the false symmetric
vacuum and the true broken vacuum so that it seems like a static
domain wall forming a boundary of two phases. It is a half-brane
describing $(p-1)$-dimensional boundary of the unstable D$p$-brane
in the middle of decay.

Array of kink-antikink and topological BPS kink have been obtained
in BCFT~\cite{Sen:2003bc},
while the latter three configurations
have not been found in BCFT, yet.
According to BSFT, EFT, and NCFT, the latter three brane configurations
are likely to exist
in BCFT too, so that they can be established
as brane configurations in superstring theory. The present BSFT analysis
opens a possibility of study including off-shell contributions.

\section*{Acknowledgements}
This work was supported by the Science Research Center Program of the
Korea Science and Engineering Foundation through the Center for
Quantum Spacetime(CQUeST) of Sogang University with grant
number R11-2005-021(C.K.), and is
the result of research activities (Astrophysical Research
Center for the Structure and Evolution of the Cosmos (ARCSEC))
supported by Korea Science $\&$ Engineering Foundation(Y.K. and O.K.).
H.U.Y. is
supported by grant No. R01-2003-000-10391-0 from the Basic
Research Program of the Korea Science \& Engineering Foundation.

\appendix
\renewcommand{\theequation}{A.\arabic{equation}}
\setcounter{equation}{0}

\section{Pressureless Limit of Array Solution}

Here we give a detailed analysis on the distance between kink and adjacent
antikink in the pressureless limit $T^{11}\rightarrow 0$
($T_{\max}\rightarrow\infty$) for pure tachyon case of section 3.

In terms of the function $G(y)$ in (\ref{G}), the expression of pressure
(\ref{t11}) is written as
\begin{equation}\label{eqg}
{{-T^{11}}\over {\cal T}_p} e^{{1\over4}T^2}
=e^{{1\over4}(T^2-T_{\rm max}^2)}=G(y).
\end{equation}
We are interested in the cases of deep potential well where
$0<-T^{11}\ll{\cal T}_p$, so that
the left-hand side of (\ref{eqg}) is much less than unity
unless $T \sim T_{\rm max}=\sqrt{\ln|{\cal T}_{p}/T^{11}|}\gg 1$.
Let us focus on the kink case, $T' >0$, in detail because an anti-kink
profile is a simple flipped copy of that of a kink.
Qualitatively speaking, when $T\ll T_{\rm max}$, (\ref{eqg}) tells us
$G(T'^2)\ll 1$ which implies that $T'\gg 1$ and the tachyon grows very fast.
As $T$ approaches $T_{\rm max}$, the left-hand side of (\ref{eqg})
goes to unity and we have $T'\to 0$ as $G(T'^2)\to 1$. Tachyon profile
turns around at the point $T=T_{\rm max}$.

More precisely, $G(T'^2)\ll 1$ when
$T \le T_{\rm max}-{2\over T_{\rm max}}\log{(T_{\rm max})}$, so that
$e^{{1\over4}(T^2-T_{\rm max}^2)}\le {1 \over T_{\rm max}}\ll 1$.
We call this kink region. On the other hand,
when $(T_{\rm max}-T)T_{\rm max} \ll 1$,
we would have $G(T'^2)\sim 1$, and for a consistent analysis,
we define the plateau region to be
$T\ge T_{\rm max}-{2\over {T_{\rm max}\log{(T_{\rm max})}}}$.
The transition region will then be
$T_{\rm max}-{2\over T_{\rm max}}\log{(T_{\rm max})}\le T\le
T_{\rm max}-{2\over {T_{\rm max}\log{(T_{\rm max})}}}$.
We therefore analyze (\ref{eqg}) in three separate regions
and we interpolate these regions. We will show that the width of each
region vanishes as $T_{\max}$ goes to infinity.

In the kink region, we can approximate
$G(T'^2)\approx {\sqrt{\pi}\over 4} {1\over T'}\ll 1$,
because $T' \gg 1$ as we have seen in the above.
Using this, (\ref{eqg}) is integrated at once to give us
\begin{equation}
\int_0^{T(x)}\, e^{{1\over 4}t^2}\, dt \,\,
=\,\, {\sqrt{\pi}\over 4}\left({\cal T}_p \over -T^{11}\right)x \,\,=\,\,
{\sqrt{\pi}\over 4}\,e^{{1\over 4} T_{\rm max}^2}\, x.\label{center}
\end{equation}
For small $T(x)$, we have
\begin{equation}
T(x)\,\,\approx\,\,{\sqrt{\pi}\over 4} \,
e^{{1\over 4}T_{\rm max}^2}\,x +{\cal O}(x^3),
\end{equation}
and this solution is valid until $T\sim {\cal O}(T_{\rm max})\gg 1$.
In addition, using
\begin{equation}
\int_0^{T}\, e^{{1\over 4}t^2}\, dt\,\,
\approx \,\,{2\over T}\, e^{{1\over 4}T^2},
\end{equation}
for large $T\gg 1$, we have
\begin{equation}
x\,\,\sim \,\,{8\over \sqrt{\pi} \,T}\,
e^{{1\over 4}(T^2-T_{\rm max}^2)},
\end{equation}
for $T\sim {\cal O}(T_{\rm max})$. Therefore, we see that the kink region
has a width of
$\Delta x\sim {8\over \sqrt{\pi}}{\left(1\over T_{\rm max}\right)^2} \ll 1$.
In other words, the tachyon
profile grows very fast up to ${\cal O}(T_{\rm max})$
in a narrow region of ${\cal O}(1/T_{\rm max}^2)$.

We next analyze the plateau region where $T$ is much close to
$T_{\rm max}$. Since the left-hand side of (\ref{eqg})
is ${\cal O}(1)$, we have $T'\sim 0$ as $G(T'^2)\to 1$
so that we can approximate $G(T'^2)\approx 1-(2\log{2})T'^2$.
Defining $\Delta T\equiv T-T_{\rm max}$ and linearizing
(\ref{eqg}) with respect to $\Delta T$, we get
\begin{equation}
\Delta T'\,\,\approx\,\, \sqrt{\left(T_{\rm max}\over 4\log{2}\right)
(-\Delta T)},
\end{equation}
which is integrated to be
\begin{equation}
\Delta T\,\,\approx\,\,- \left(T_{\rm max} \over 16\log{2}\right)
(x-x_0)^2 ,\label{nearm}
\end{equation}
where $x_0$ is the turning point of the tachyon profile.
The width of this region is easily calculated by letting
$\Delta T=-{2\over T_{\rm max}}{1\over \log{(T_{\rm max})}}$
to get $\Delta x = \sqrt{32\log{2}\over \log{(T_{\rm max})}}
\, {1 \over T{\rm max}}$,
which becomes zero as $T_{\rm max}\to \infty$.
Note that the gradient of tachyon profile in (\ref{nearm})
is very steep as $T_{\rm max}\to \infty$ and this profile crosses
over to a even steeper profile (\ref{center})
of the kink region.

The analysis of the transition region turns out to be much more subtle,
especially the width of the region.
Defining $u\equiv T_{\rm max}(T_{\rm max}-T)$, the equation (\ref{eqg})
approximates well to
\begin{equation}
G(u'^2 / T_{\rm max}^2)\,\,=\,\, e^{-{u\over 2}}.
\end{equation}
Through a tedious but straightforward analysis of the above equation,
we find that the width $\Delta x$ is of order ${\cal O}(1/ T_{\rm max})$,
so that this dominates over those of the other two regions.
This establishes that the whole width of a single kink becomes zero
as $T_{\rm max}\to \infty$.

Now let us estimate the energy.
Looking at the energy density (\ref{ede}), and
${\cal F}(x)\approx \sqrt{\pi x}+{\cal O}(x^{-{1\over 2}})$ for $x\gg 1$,
we see that the energy density in the kink region is
\begin{equation}
T^{00}\,\,\approx\,\,\sqrt{\pi}{\cal T}_p \,e^{-{1\over 4}T^2}\,|T'|\,\,
\approx\,\,{\pi\over 4}{\cal T}_p \,
e^{-{1\over 2}T^2+{1\over 4}T_{\rm max}^2},
\end{equation}
which is highly peaked near $T=0$.
Moreover, the solution (\ref{nearm}) around $T\sim T_{\rm max}$
shows that $|T'| \gg 1$ unless $(x-x_0) \sim {\cal O}(1/T_{\rm max})$,
which in turn corresponds to $\Delta T \sim {\cal O}(1/T_{\rm max})$.
This means that the validity of the approximation
$ {\cal F}(T'^2)\approx \sqrt{\pi}\,T'$ we have used before extends up
to $T=T_{\rm max}-{\cal O}(1/T_{\rm max})$.
Therefore,
the energy integral for a kink approximates to
\begin{equation}
\int \, T^{00}\,dx \,\,\sim\,\,\sqrt{\pi}{\cal T}_p
\int\,e^{-{1\over 4}T^2}\,T'\,dx\,\,
=\,\,\sqrt{\pi}{\cal T}_p\int\,e^{-{1\over 4}T^2}\,dT,
\end{equation}
with the integration region
$-T_{\rm max}+{\cal O}(1/T_{\max})\le T \le T_{\max}-{\cal O}(1/T_{\max})$.
Using ${\cal F}(x)\approx 1+{\cal O}(x)$ for $x\ll 1$, the contribution
from the region $\Delta T \le {\cal O}(1/T_{\max})$ can be shown to be
of order
${\cal O}\left(\frac{{\cal T}_p}{T_{\rm max}}
e^{-{1\over 4} T_{\rm max}^2}\right)$,
which is negligible.
Hence, the topological nature of a kink(anti-kink) energy becomes
extremely good as $T_{\rm max}\to \infty$ ($T^{11}\to 0$),
and its value goes to
\begin{equation}
{\cal T}_{p-1}=\sqrt{\pi}{\cal T}_p\,
\int_{-\infty}^{\infty}\,\,e^{-{1\over 4}T^2}\,dT
=2\pi\,{\cal T}_p =(2\pi\sqrt{\alpha'} )\,{{\cal T}_p\over \sqrt{2}}.
\end{equation}
This establishes (\ref{puret}).

\section{Comparison with Bosonic Boundary String Field Theory}

\begin{figure}[ht]
\begin{center}
\scalebox{0.55}[0.55]{
\includegraphics{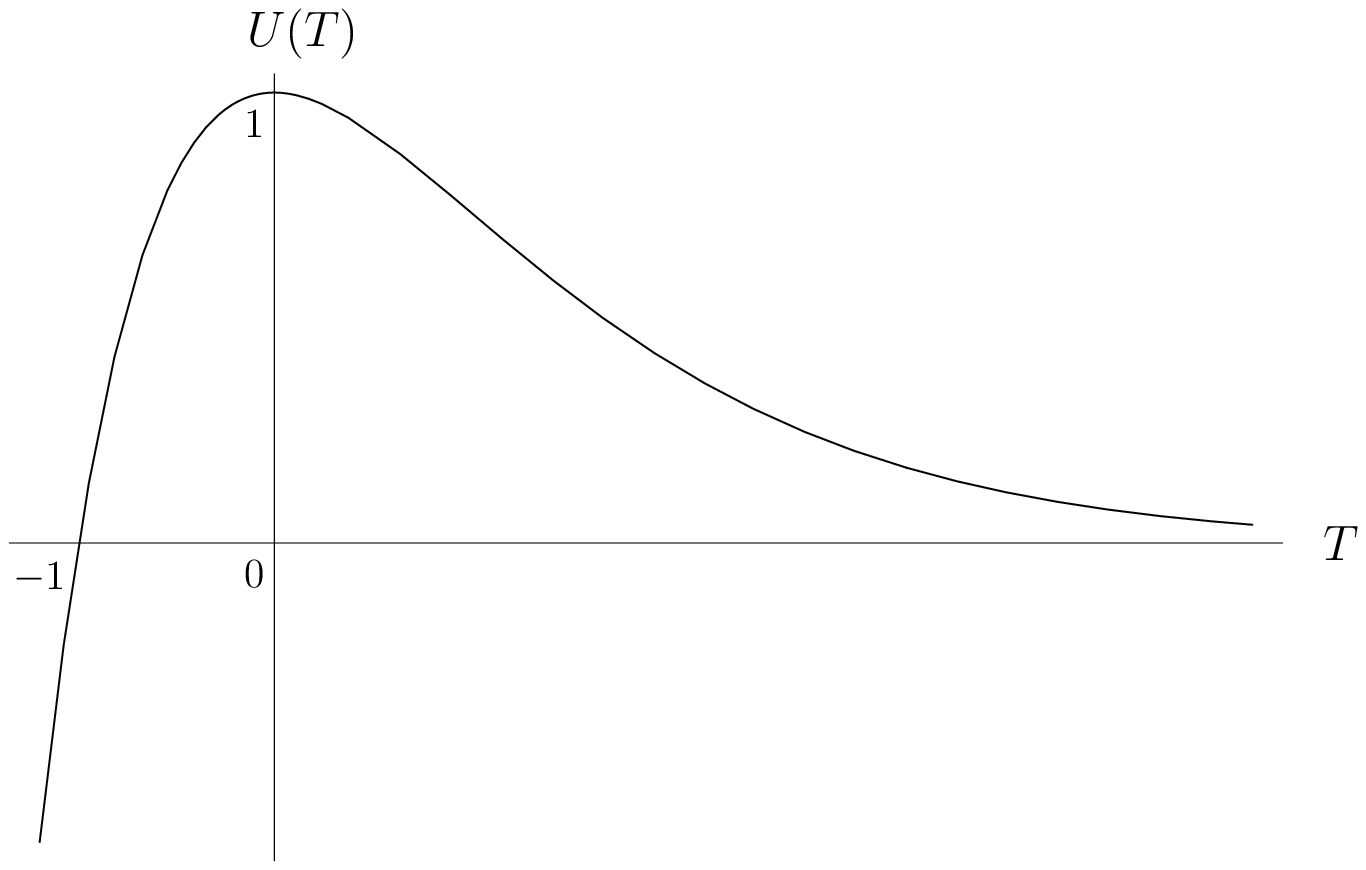}
\includegraphics{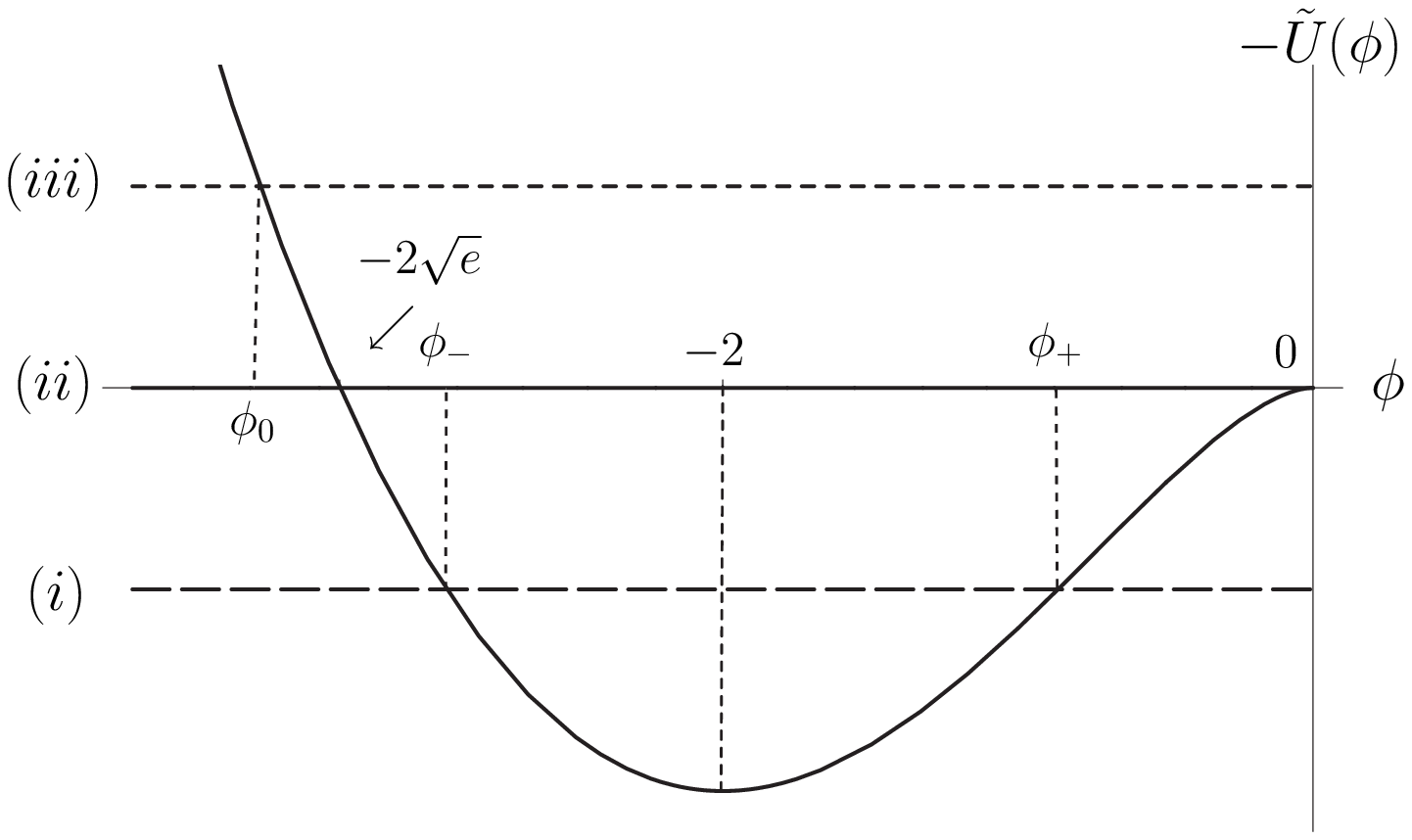}}
\par
\vskip-2.0cm{}
\end{center}
\caption{\small The graphs of $K(x^2)$(left) and $U(x)$(right).
For $U(x)$, three cases of $\left({{\cal T}_p \over -T^{11}}\right)
={1\over \sqrt{2}},1,\sqrt{2}$ from top to bottom are shown. }
\label{figa}
\end{figure}

In this appendix we look into possible tachyon kink solutions in bosonic
BSFT~\cite{Gerasimov:2000zp}.
Differently from the super BSFT, the tachyon potential is exact but
its kinetic term is approximate because of impossibility of exact
computation of the worldsheet beta function.
When higher-derivative terms are neglected, the tachyon action is given as
\begin{equation}
S_{\rm B} = -{\cal T}_p\int d^{p+1} x \left[ 2 e^{-T}
\partial_\mu T \partial^\mu T + ( T + 1) e^{- T} + \cdots\right],
\end{equation}
where the exact tachyon potential involves instability of the bosonic
string theory for negative $T$ and that of the unstable D$p$-brane for
positive $T$ as shown in Fig.~\ref{figa} (left).

A field redefinition of the tachyon $\phi = -2 e^{-\frac{T}{2}}$ lets the
quadratic derivative term have a canonical form,
\begin{equation}
S_{{\rm B}} = -{\cal T}_p\int d^{p+1} x \left[
2\partial_\mu\phi \partial^\mu\phi + \tilde U(\phi)\right],
\end{equation}
and then the shape of tachyon
potential as a function of the redefined tachyon field $\phi$,
\begin{equation}
\tilde U(\phi) = \left[1- \ln
\left(\frac{\phi}{2}\right)^2\right]\left(\frac{\phi}{2}\right)^2 ,
\end{equation}
resembles that from vacuum string field theory~[REF],
i.e., the distance
between the local maximum at $\phi=-2$ and the local minimum at $\phi=0$
is finite.

Under the static ansatz of the tachyon kink $\phi=\phi(x)$, the following
first-order equation is obtained
\begin{equation}\label{a1}
{\tilde {\cal E}}={\dot \phi}^{2}
+\frac{-{\tilde U}}{2}.
\end{equation}
According to value of the integration constant ${\tilde {\cal E}}$,
analysis can be made by dividing the cases into three (see
the dashed line of (i), the solid line of (ii), and the dotted line
(iii) in Fig.~\ref{figa} (right)).
When ${\tilde {\cal E}}>0$ (see (iii)),
absence of the potential for positive $\phi$
does not allow a proper kink configuration which spans whole range of $x$.
Array of kink and antikink for $-1<{\tilde {\cal E}}<0$ (see (i))
is likely to be a natural solution consistent with that in the super BSFT
in section 3. For the critical value ${\tilde {\cal E}}=0$, there
seems to exist a pair of kink and antikink, but its existence should be
confirmed by further study including all the higher derivative
terms.

\end{document}